\documentclass[letterpaper,twocolumn,10pt]{article}
\usepackage{usenix}
\usepackage{hyperref}
\usepackage[hyphenbreaks]{breakurl}
\usepackage{amssymb}
\usepackage{filecontents}
\usepackage{balance}
\usepackage{amsmath, amsthm}
\usepackage{mathtools, nccmath}
\usepackage{url}
\usepackage{color}
\usepackage{caption}
\usepackage{diagbox}
\usepackage{subcaption}
\usepackage{multirow}
\usepackage{booktabs}
\usepackage{tikz}
\usepackage{epsfig,endnotes}
\usepackage{times}
\usepackage{algorithm,algpseudocode}
\usepackage{tablefootnote}
\usepackage{booktabs}

\newcommand\numberthis{\addtocounter{equation}{1}\tag{\theequation}}
\newtheorem{theorem}{Theorem}
\newtheorem{definition}{Definition}
\definecolor{ocean}{RGB}{78, 168, 245}

\newcommand{\mnist}{{\tt MNIST}}
\newcommand{\gtsrb}{{\tt GTSRB}}
\newcommand{\cifar}{{\tt CIFAR10}}
\newcommand{\imagenet}{{\tt ImageNet}}
\newcommand{\imdb}{{\tt IMDB}}

\newcommand{\para}[1]{{\vspace{1.2pt} \noindent \textbf{#1}
    \hspace{6pt}}}

\newcommand{\spara}[1]{{\vspace{0.5pt} {\em #1}
    \hspace{3pt}}}

\definecolor{ForestGreen}{RGB}{24,155,24}
\newcommand{\huiying}[1]{{\color{black} #1}}

\newcommand{\htedit}[1]{{\color{black} #1}}

\newcommand{\etal}{{\em et al.\ }}
\newcommand{\eg}{{\em e.g.,\ }}

\newenvironment{packed_itemize}{
\begin{list}{\labelitemi}{\leftmargin=0.8em}
  \setlength{\itemsep}{1pt}
  \setlength{\parskip}{0pt}
  \setlength{\parsep}{0pt}
  \setlength{\headsep}{0pt}
  \setlength{\topskip}{0pt}
  \setlength{\topmargin}{0pt}
  \setlength{\topsep}{0pt}
  \setlength{\partopsep}{0pt}
}{\end{list}}

\begin{document}

\title{Blacklight: Scalable Defense for Neural Networks against \\ Query-Based Black-Box Attacks}

\author {Huiying Li, Shawn Shan, Emily Wenger, Jiayun Zhang, Haitao
Zheng, Ben Y. Zhao\\
{\em Computer Science, University of Chicago}\\
{\em \{huiyingli, shansixiong, ewillson, jiayunz, htzheng, ravenben\}@cs.uchicago.edu}}
\maketitle

\begin{abstract}



\htedit{ Deep learning systems are known to be vulnerable to adversarial
examples. In particular, query-based black-box attacks do not require
knowledge of the deep learning model, but can compute adversarial
examples over the network by submitting queries and inspecting
returns. Recent work largely improves the efficiency of those attacks,
demonstrating their practicality on today's ML-as-a-service
platforms.}

We propose {\em Blacklight}, a new defense against query-based
black-box adversarial attacks. The fundamental insight driving
our design is that, to compute adversarial examples, these
attacks perform iterative optimization over the network, producing
image queries highly similar in the input space.
Blacklight detects query-based black-box attacks by detecting
highly similar queries,  using an efficient similarity engine operating on probabilistic content fingerprints.
We evaluate Blacklight \htedit{against} eight state-of-the-art
attacks, across a variety of models and image classification
tasks.  Blacklight identifies them all, often after only a handful of
queries. \htedit{By rejecting all detected queries,  Blacklight
  prevents any attack to complete, even when attackers persist to submit queries
  after account ban or query rejection.} Blacklight is also robust against several powerful countermeasures,
  including an optimal black-box attack that approximates white-box attacks
  in efficiency. \htedit{Finally, we illustrate how Blacklight generalizes to
other domains like text classification. }


\end{abstract}

\vspace{-0.1in}
\section{Introduction}
\vspace{-0.05in}
\label{sec:intro}

The vulnerability of deep neural networks (DNNs) to a variety of adversarial
examples is well documented.  An adversarial example is a maliciously
modified input that looks (nearly) identical to its original via
human perception, but gets misclassified by a DNN model.
This vulnerability remains a
critical hurdle to the practical deployment of deep learning systems in
safety- and mission-critical applications, such as autonomous driving or
financial services. 

Adversarial attacks can be broadly divided by whether they assume {\em
  white-box} or {\em black-box} threat models. In the {\em white-box} setting, the
attacker has total access to the target model, including its internal
architecture, weights and parameters. Given a benign input, the attacker can
directly compute adversarial examples as an optimization problem. In contrast, an
attacker in the {\em black-box} setting can only interact with the model
by submitting queries and inspecting returns.
Black-box scenarios can be
further divided based on the information the classifier returns per
query: {\em score-based} systems return a full probability distribution across
labels, and {\em decision-based} systems return only the output label.

The white-box threat model makes a strong assumption: an attacker has
obtained total access to the model, through a server breach, a malicious
insider, or other type of model leak. Both security and ML communities have
made continual advances in both attacks and defenses under this
setting -- powerful attacks efficiently generate adversarial
examples~\cite{uesato2018adversarial,carlini2017towards,chen2018ead,kurakin2016adversarial,goodfellow2014explaining},
which in turn spur work on robust defenses that either prevent the generation
of adversarial examples or detect them at inference time.  While numerous
approaches have been explored as defenses (e.g., model
distillation~\cite{papernotdistillation}, gradient
obfuscation~\cite{buckman,dhillon,ma2018characterizing,samangouei,song,xie},
adversarial training~\cite{zheng2016improving, madry2017towards,
  zantedeschi2017efficient}, and ensemble
methods~\cite{tramer2018ensemble}),
nearly all have been proven vulnerable to followup
attacks~\cite{distallationbroken,bypassten,ensemblebroken,magnetbroken,obfuscatedicml}.

 \begin{figure}[t]
    \centering    \vspace{-0.1in}
    \includegraphics[width=.36\textwidth]{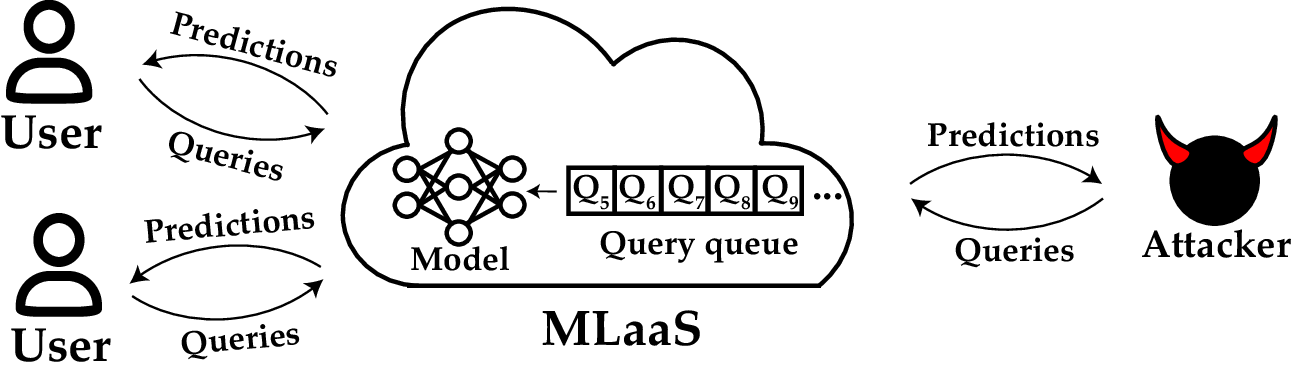}
      \vspace{-0.06in}
      \caption{\em Attack Scenario for black-box adversarial attacks.}
      \label{fig:attack_scenario}
       \vspace{-0.2in}
  \end{figure}

In contrast, black-box attacks assume a more realistic threat model, where
attackers interact with models via a query interface such as ML-as-a-service
platforms~\cite{mlaas} (See Fig~\ref{fig:attack_scenario}).  There are two
types of black-box attacks. Most common are
{\em query-based attacks}~\cite{chen2017zoo, ilyas2018black, bhagoji2018practical,
  moon2019parsimonious, tu2019autozoom, chen2020hopskipjumpattack}, where an
attacker iteratively adapts the query input based on past query results from
the target model, until it produces a successful adversarial
example. Numerous efforts have developed increasingly efficient
attacks \htedit{that require fewer queries to complete the attack}.
Unfortunately, even as these
attacks grow in efficiency and practicality,
\htedit{there exists no effective defense against them. Existing
  defense proposals~\cite{chen2020stateful,
    juuti2019prada} focus on detecting (and banning) query accounts
  displaying some
  ``adversarial'' behaviors. While raising the attack
  cost, they are ineffective against persistent attackers
  who switch accounts to evade detection and complete the attack.}
The second type of black-box attacks is
   {\em substitute model attacks}, where an attacker
queries the target model to train a local model, then tries
to transfer adversarial examples from the substitute to the
target~\cite{liu2017delving,papernot2016transferability,papernot2017practical}.
These are currently addressed by a line of effective and
  evolving defenses, including (ensemble) adversarial
  training~\cite{tramer2018ensemble,wong2020fast}.


\htedit{In this work, we focus on defending against query-based black-box
    attacks, even when persistent attackers switch account to evade detection.}
The fundamental insight driving our work is that, in order to compute
adversarial examples, query-based black-box attacks {\em perform iterative
  optimization over the network}, an incremental process that produces
queries highly similar in the input space. \htedit{With this in mind}, we
propose {\em Blacklight}, a novel defense that detects query-based black-box
attacks using \htedit{an efficient {\em content-similarity engine}.}
  Blacklight detects the highly similar
  queries
  as part of the iterative optimization process in the
attack\footnote{In practice, even the most efficient black box attacks issue
  thousands of queries to generate a single attack, and nearly all such
  queries are constrained to be a small perturbation away from the benign
  input.}, since benign queries rarely share this level of
similarity. \htedit{Blacklight's query detection is account
  oblivious, thus is effective no matter how many accounts an
  attacker uses to submit queries.}




\htedit{Blacklight is highly scalable and lightweight. It} detects highly similar queries generated by
iterative
optimization using {\em probabilistic fingerprints},
\htedit{a compact hash representation computed for each input
query.  We design these fingerprints such that queries highly similar
in the input space
will have large overlap in their fingerprints.
As such, Blacklight identifies an (incoming) query as part of a query-based black-box
attack, if its fingerprint matches any prior fingerprint by more than
a threshold. }
 Since we use
secure one-way hashes to compute fingerprints, even an attacker aware of our algorithm cannot
optimize \htedit{the content perturbation of a query} to disrupt its fingerprint and avoid detection.


We evaluate the efficacy of Blacklight against
eight  SOTA query-based
black-box attacks, including those using gradient
  estimation,  gradient-free attacks, and those targeting score- and decision-based models.
We experiment on a range of image-based
models from MNIST to ImageNet, and use $L_p$ distance metrics chosen by each
attack. While these attacks typically take thousands
(or tens of thousands) of queries to converge to a successful adversarial
example, Blacklight detects all of them after the first 2--9
queries\footnote{The exception is the Boundary attack, which starts its query
search with an image from the target label. Blacklight detects Boundary attacks after an
average of less than 50 queries (see Table~\ref{table:results}).}.
More importantly, Blacklight detects the
large majority of all queries associated with an attack (e.g., $>$90\% for all
non-Boundary attacks).  \htedit{By rejecting these detected attack queries,
Blacklight consistently reduces the attack success rate to 0\% for all
eight attacks, even when attackers
persist to submit queries despite query rejection.


}

Our work makes the following key contributions.
\begin{packed_itemize} \vspace{-0.1in}
  \item We propose a highly scalable, lightweight attack detection system against
    query-based black-box attacks,  using \htedit{probabilistic
      content fingerprint-based query matching to detect (and mitigate) individual attack
      query on the fly.}
   \item \htedit{We discuss and demonstrate why existing account-based
  defenses are insufficient to resist persistent attackers. }
  \item We build formal analysis of our probabilistic fingerprints to model both attack detection
    rates and false positives.
  \item We experimentally evaluate Blacklight against eight SOTA
    black-box attacks on multiple datasets and \htedit{image classification} models. Not only does
    Blacklight detect all eight attacks, but it does so {\em quickly}, often after
    only a handful of queries, for attacks that would require several
    thousands of queries to succeed.

\item \htedit{We illustrate how Blacklight can be generalized beyond
    image classification, using text classification as an example.}
  \item We finally evaluate Blacklight and show it is highly robust
    against a variety of adaptive countermeasures, including those allowing
    larger, human-visible perturbations. Blacklight performs well even
    against two types of {\em near-optimal} attacks:
    ``query-efficient''  attacks several orders of
    magnitude more efficient than current methods, and  ``perfect-gradient''
    attacks that approximate white-box attacks by perfectly
    estimating the loss surface at each query.



    \vspace{-0.1in}
  \end{packed_itemize}

  \huiying{
  The source-code for Blacklight is at
      \url{https://github.com/Huiying-Li/blacklight}. }

\vspace{-0.15in}
\section{Background on Black-box Attacks}
\label{sec:back}
\vspace{-0.05in}


As background, we briefly overview different types of black-box
attacks and describe today's SOTA query-based black-box attacks (the focus
of our work). \htedit{We discuss existing defense
  proposals~\cite{chen2020stateful, juuti2019prada} later in
  \S\ref{sec:existing}. }






\vspace{-0.15in}
\subsection{Overview of Black-box Attacks}
\label{sec:backoverview}
\vspace{-0.05in}
Existing black-box attacks can be divided into two
types: substitute model attacks and
query-based black-box attacks.  In this work, we target the latter.

\para{Substitute Model Attacks.}  An attacker \htedit{queries a target
  model repeatedly},  
  uses the query results to build a
labeled dataset and train a {\em substitute} model to approximate
classification boundaries of the model.  The attacker then generates
adversarial examples on the substitute model (using a white-box
attack), hoping that they will succeed on the target model.
This attack is shown to successfully produce untargeted
adversarial examples on small models like
MNIST~\cite{papernot2016transferability, papernot2017practical}, but
\htedit{become much less successful when}
  producing targeted attacks or going against
larger models~\cite{liu2017delving}. This spurs efforts to
increase
transferability between substitute and target models~\cite{xie2019improving,
  dong2019evading, wu2020skip, inkawhich2020perturbing, lin2019nesterov}.

Defending against substitute model attacks is an active
research area.  Existing defenses include adversarial
training~\cite{kurakin2017adversarial}, ensemble adversarial
training~\cite{tramer2018ensemble}, and adversarial training with single-step
R+FGSM attack~\cite{wong2020fast}.
We note that ensemble adversarial training can be combined with Blacklight as
a hybrid defense against both substitute model attacks and query-based
attacks (details in the Appendix \S\ref{sec:hybrid_submodel}).

\label{sec:back_query}
\para{Query-Based Black-Box Attacks.}  A more common and effective
attack is query-based black-box attacks. An attacker queries the target model
repeatedly, often remotely over a network, to implement iterative
optimization required to compute adversarial examples. Specifically, based on the past query results, the attacker
iteratively perturbs the current query to produce the next query, hoping to
converge to a successful adversarial example.    \htedit{Both gradient-estimation~\cite{chen2017zoo, ilyas2018black,
  tu2019autozoom, chen2020hopskipjumpattack, cheng2019sign} and gradient-free algorithms~\cite{bhagoji2018practical,
  moon2019parsimonious, andriushchenko2020square} were developed to reduce the number of queries required to produce an adversarial example.}
While these attacks generally
require thousands to hundreds of thousands of queries to produce a single
adversarial example, they have proven to be effective, often achieving 100\%
success rate even against large models.  In fact, recent efforts
show that these attacks can already be successfully launched against
real-world systems such as Google Cloud Vision API~\cite{ilyas2018black},
Clarifai~\cite{bhagoji2018practical}, and real applications like traffic sign
and license plate recognition~\cite{feng2020query}.  \htedit{Finally,
   recent works also leverage substitute model-based priors when configuring queries~\cite{suya2020hybrid,
    juuti2019making, huang2019black, cheng2019improving}, which we
  also consider when evaluating Blacklight in \S\ref{subsec:reduce_queries}.}

\vspace{-0.1in}
\subsection{SOTA Query-based Black-box Attacks}
\vspace{-0.05in}
\label{sec:selected_attacks}
Our work targets query-based black-box
attacks. We implement and test eight SOTA attacks (see Table~\ref{table:attack_category}). They cover both
score- and decision-based attacks, and attacks relying on
gradient estimation and those that do not.  They all use $L_p$ bounded
perturbations, a  prevailing
attack setting.

\begin{table}[h]
\centering  \vspace{-.1in}
\resizebox{.47\textwidth}{!}{
\begin{tabular}{ c|c|c}
\hline
 & \multicolumn{1}{l|}{{\bf Gradient Estimation}} &
                                                           \multicolumn{1}{l}{{\bf
                                                          Gradient Estimation Free}} \\ \hline
{\bf Score-based} & NES - Query Limit\cite{ilyas2018black} & \begin{tabular}[c]{@{}c@{}}ECO\cite{moon2019parsimonious}\end{tabular} \\ \hline
{\bf Decision-based} & \begin{tabular}[c]{@{}c@{}}NES - Label
                         Only\cite{ilyas2018black}\\
                         HSJA\cite{chen2020hopskipjumpattack} ~~  QEBA\cite{li2020qeba} \\Policy-Driven\cite{yanpolicy}\end{tabular} & \begin{tabular}[c]{@{}c@{}}Boundary\cite{brendel2018decision} \\SurFree\cite{maho2020surfree}\end{tabular} \\  \hline
\end{tabular}}
 \vspace{-.1in}
 \caption{\em We consider eight query-based  black-box attacks.}
 \label{table:attack_category}
 \vspace{-0.15in}
\end{table}

\para{NES (2 variants)~\cite{ilyas2018black}.}  NES enables efficient gradient
estimation using far fewer queries and applies natural
evolution strategies~\cite{wierstra2008natural} to speed up the attack.
NES has
two variants: {\em NES query limit} for score-based models and {\em
  NES label-only} for decision-based models.


\para{ECO~\cite{moon2019parsimonious}.}  Targeting score-based models,
the attacker replaces gradient
estimation with an efficient discrete surrogate, leading to faster convergence.

\para{Boundary~\cite{brendel2018decision}.} It is the \htedit{first
  attack targeting decision-based models and does not use gradient
estimation. }
To compute the adversarial example for $x_0$,
the attacker starts from a random sample $x$ from the target label
$t$,  iteratively adjusts $x$ to ``approach'' $x_0$ while remaining
being classified to $t$, until the difference between $x_0$ and $x$ is within a predefined budget.


\para{HSJA~\cite{chen2020hopskipjumpattack}.} It augments Boundary~\cite{brendel2018decision} with gradient
approximation. In each iteration, a 2-step gradient estimation is used
to construct $x_t$ that gets closer to the decision boundary, leading to
much faster attack convergence than Boundary.


\para{QEBA~\cite{li2020qeba}.} This is a variant of HSJA. Instead of
estimating the full gradient vector,  QEBA
only estimates a core subset of the gradient vector.

\para{Policy-Driven~\cite{yanpolicy}).} This is another recent attack
built
on top of HSJA. It applies a policy network to \emph{learn} the best
optimization direction at each step.

\para{SurFree~\cite{maho2020surfree}.} This gradient-free attack
leverages certain geometrical properties to  produce careful query trials along diverse directions
near the decision boundaries.


\vspace{-0.1in}
\section{Threat Model and Design Goals}
\label{sec:threat}
\vspace{-0.06in}
In this work, we focus on defense against query-based black-box attacks for image
classification.  \htedit{Our design principle should extend to other
domains, which we demonstrate in \S\ref{sec:other_domain} using text
classification as an example.}
Here, we define our threat model,  design goals and success
metrics.

\para{Attacker.} The attacker queries a target DNN model
($\mathbb{F}$) and uses the query results to craft adversarial examples
against it, i.e.,
finding the perturbed version of a benign input $x_0$ that causes  $\mathbb{F}$
to {\em misclassify} it to a target label $t$.  To do so, the
attacker repeatedly queries $\mathbb{F}$ with a sequence of $n$ attack
queries $x_1, ..., x_n$ (i.e.,  started from $x_0$ and ended with
$x_n$).  The attack is successful if
\begin{equation} \vspace{-0.04in}
  \mathbb{F}(x_n) = t \;\; \text{ and }\;\;||x_n - x_0||_p < \epsilon
  \label{eq:attackperturbation}\vspace{-0.0in}
\end{equation}
where $x_n$ is the computed adversarial example of $x_0$ and $\epsilon$ is the
attacker's perturbation budget. Existing works show that a successful attack
requires a large $n$, generally on the order of $10^3$-$10^6$.
Note that while we focus on prevailing attacks that bound perturbations by
  L$_p$ distance, our defense should extend in principle to other query-based
  attacks (\eg patch, semantic attack).  We discuss in \S\ref{subsec:patchresult} preliminary results on Sparse-RS~\cite{croce2020sparse}, a query-based
  universal patch attack.


We make the following assumptions about the attacker:
\begin{packed_itemize} \vspace{-0.08in}
\item The attacker has no access to internal weights of
  $\mathbb{F}$ and can only send queries to obtain outputs of $\mathbb{F}$.
\item The attacker has abundant computation power and resources to submit
  millions of queries.
\item The attacker controls {\bf multiple} user accounts and IP addresses, and
  moves the attack across them if any IP addresses and/or accounts are banned.
  Measurements have shown that attackers often utilize
  {\em Sybil} accounts~\cite{sybil,yang2014uncovering,sybil-normal}.

\item  We begin with standard attackers who are unaware of
  Blacklight. Later in \S\ref{sec:countermeasure}, we consider stronger
  adaptive attackers who apply countermeasures against
  Blacklight.   \vspace{-0.06in}
\end{packed_itemize}

\para{Defender.} The defender hosts the target model $\mathbb{F}$.
For each query, $\mathbb{F}$ can either return the full classification
probability vector or only the classification label. We only make one assumption on the defender, that it has a {\bf finite} amount of
storage for use in attack detection. In practical terms, any defender
storing state related to past queries has to periodically \textbf{reset}
the storage, e.g., every 1 or 2 days, \htedit{by clearing out the state of
all past (benign) queries}.

\para{Design Goals.} We target four key goals for our defense.
\begin{packed_itemize}
  \vspace{-0.05in}
\item The defense should detect attack queries with \textbf{high
  accuracy} and {\bf high coverage}, while maintaining a \textbf{low false positive
  rate}. \htedit{Since answering each attack query may leak model
  information, the defense should detect as many attack queries as possible.}

\item The defense should efficiently \textbf{scale} to industry production
  systems. For example, Facebook's content moderation systems process an
  average of 300M images per day, while those at Twitter process 340M
  tweets/day~\cite{fbscale,twitterscale}.
\item The defense should incur \textbf{low  overhead} \htedit{in
    terms of
runtime (compared to model inference runtime) and storage}.

\item \htedit{The defense must {\bf resist persistent attackers} who can
    move between accounts, and/or continue
    submitting attack queries after account ban or query rejection.}
\vspace{-0.1in}
\end{packed_itemize}




\vspace{-0.13in}
\section{Existing Defenses and Their Limitations}
\vspace{-0.08in}
\label{sec:existing}

  \htedit{There are two known defenses
  against query-based black-box attacks: \textit{Stateful Detection}~(SD)~\cite{chen2020stateful}
  and PRADA~\cite{juuti2019prada}. Both are
  account-driven and focus on detecting/banning query
  accounts that submit
  attack queries. We now describe their
  detection methods, and discuss why these
  defenses (and their variations) are insufficient to resist
  persistent attackers covered by our threat model.

   \para{Stateful Detection (SD)~\cite{chen2020stateful}.}  SD inspects
  each query account to decide whether it is malicious or not.
  Given an account $A$ and its queries submitted so far,  SD examines whether these queries display ``certain properties'' related
  to computation of adversarial examples. Specifically,  SD computes, for an incoming query $q$ from $A$, the
  average pair-wise latent similarity between $q$ and its
  k-nearest-neighbors in $A$'s past queries.  If the average latent
  similarity exceeds a threshold, SD flags $A$ as
  adversarial. To compute the latent similarity, SD uses a pretrained similarity encoder to convert each query image into a
  latent space vector.

  \para{PRADA~\cite{juuti2019prada}.} Originally designed to
  detect attacks that steal the
  target model, PRADA is shown to also detect query-based black-box
  attacks~\cite{chen2020stateful}. The key insight is
  that queries sent by an attacker are expected to have a
  characteristic distribution different from those of benign
  accounts. PRADA calculates the query
  distribution of each account based on the $L_2$ distance among queries, and
defines a standard benign distribution computed from a set of benign
  queries.  If an account $A$'s query distribution shifts away from
  the standard benign distribution, PRADA labels $A$ as malicious.

  }

  \htedit{ \para{Vulnerability to Persistent Attacks.} While SD and PRADA
    could flag an attacker who use a single account to send attack queries,
    they are ineffective against attackers holding multiple accounts,
    e.g. Sybil accounts~\cite{sybil}. Use of Sybil attacks by bad actors have
    been long observed in measurements of online
    systems~\cite{yang2014uncovering,wang2013you}. Figure~\ref{fig:sdtrace}
    plots an example where an attacker completes an attack, by switching
    accounts and continuing its queries after each detection event by SD.
    A similar strategy would also succeed against PRADA.

The two existing defenses are limited by two factors.
 {\em First},  inspecting queries per-account puts a fundamental limit
 on detection speed, i.e., the number of attack queries answered by
 the model before detection.  For both defenses, at the time of detection,  the attacker already had tens or more attack queries answered by the
  model (e.g., 52 - 54 queries for SD and 111-115 queries for PRADA,  per our experiments in Appendix Table~\ref{table:compare}).    {\em Second}, both
  defenses are designed to ``slow down'' attackers by banning their
  current account rather than preventing the attack query to proceed.
  Given the low cost and prevalence of sybil accounts, attackers can
  easily bypass these defenses.   A ``reactive'' strategy  is shown in
  Figure~\ref{fig:sdtrace}, where 6 out of 328 attack
  queries (or 1.8\%) were detected and rejected
        and 322 got answered.   An alternative ``proactive'' strategy is to
  first run test cases to estimate the minimum \# of attack queries to get the
  account banned (e.g., 50), and then during the attack,  send
  less queries per account (e.g., 30) to evade detection completely.

 \begin{figure}[t]
    \centering
    \includegraphics[width=.4\textwidth]{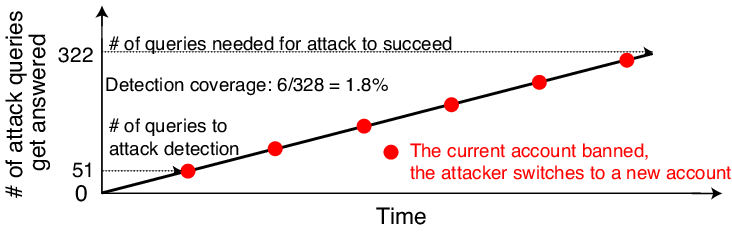}
      \vspace{-0.1in}
      \caption{\em Existing defenses cannot stop persistent attackers
        who switch accounts to continue attack queries. }
      \label{fig:sdtrace}
       \vspace{-0.15in}
  \end{figure}

  \para{Adapting Account-based Defenses.} Account-based query inspection and mitigation is ineffective
  against attackers with multiple query accounts.  An effective
  defense needs to be account oblivious.  One straightforward solution is to run a version of SD or PRADA by
  putting all the queries into a single account. This solution,
  however, does not scale to support production ML systems facing
  millions of queries per day, because both SD and PRADA's runtime
  complexity grows with the number of prior queries.
Consider a query database of 1 million low-resolution images
(\cifar, $32 \times 32$), our experiments show that,  for each
incoming query,  SD and PRADA introduce a
run-time latency of  24,000\% and 6,800\% compared to the normal
inference latency, respectively (details in
\S\ref{sec:overhead}). Furthermore, PRADA faces large accuracy
drop, because each incoming query produces little impact
on the query distribution.

}

\vspace{-0.12in}
\section{Blacklight}
\label{sec:overview}
\vspace{-0.08in}

We propose \textit{Blacklight}, a new defense to detect and mitigate query-based
black-box attacks against DNN models. \htedit{Different from existing
  defenses, Blacklight is account oblivious and focuses on detecting
  individual attack queries on the fly regardless of who sent them}.  Our design is driven by
  a fundamental insight that query-based black-box attacks produce queries that
  are highly similar in the input space. Since benign queries
  rarely share this level of similarity,  these
  attacks can be detected by identifying extremely
  high similarity in queries while incurring low false positives.  With this in mind, we design Blacklight to focus on achieving fast,
scalable and robust similarity check across millions of image queries.
Our design includes two key components: (i) {\em probabilistic content
fingerprinting} for  fast and scalable attack detection, and (ii) {\em
salted pixel quantization} to resist adaptive attacks.

In the
following, we present the fundamental insight driving our design, and the concept of probabilistic content fingerprinting.
Later in \S\ref{sec:design}, we describe the salted pixel quantization and
Blacklight's detailed design.

\vspace{-0.1in}
\subsection{Fundamental Insight: Presence of High Similarity in Attack
  Queries}\vspace{-0.08in}
  \label{sec:insight}
Blacklight exploits a fundamental insight on
  query-based black-box attacks:
in order to compute adversarial
  examples, attackers need to perform iterative
  optimization {\em over the network}, i.e.,  submitting one or more queries to the
  target model,  observing the query results, and using them to
  configure
  further queries.   While the specific design of iterative
optimization is algorithm-dependent\footnote{Some attack designs start
with the original input and perturbs it towards a misclassified target
label~\cite{ilyas2018black, moon2019parsimonious}, while others start from an image in the target label and work
backwards towards the original input~\cite{ilyas2018black,
  brendel2018decision, chen2020hopskipjumpattack}.},  the unified goal
is to repeatedly refine the perturbation such that the query
sequence converges to an adversarial example $x_n$ satisfying
eq~(\ref{eq:attackperturbation}). Therefore, iterative optimization
inevitably produces {\em some} queries that are highly similar in the
input space, i.e.,
\begin{equation*}\vspace{-0.06in}
\htedit{\text{there exist}}\;\; x_k, x_j, \text{where } \,\, ||x_k - x_j||_p \leq
  \mu.  \vspace{-0.0in}
  \end{equation*}
If $\mu$ is sufficiently smaller than
the difference between most benign images,  we can accurately
detect the attack by recognizing the presence of highly similar
queries like ($x_k, x_j$) within the stream of queries.  Evading this
type of detection is extremely difficult (if not infeasible) since
it requires {\em every attack query to be sufficiently dissimilar from
  any previous attack queries}.

\begin{figure}[t]
  \centering
  \includegraphics[width=.42\textwidth]{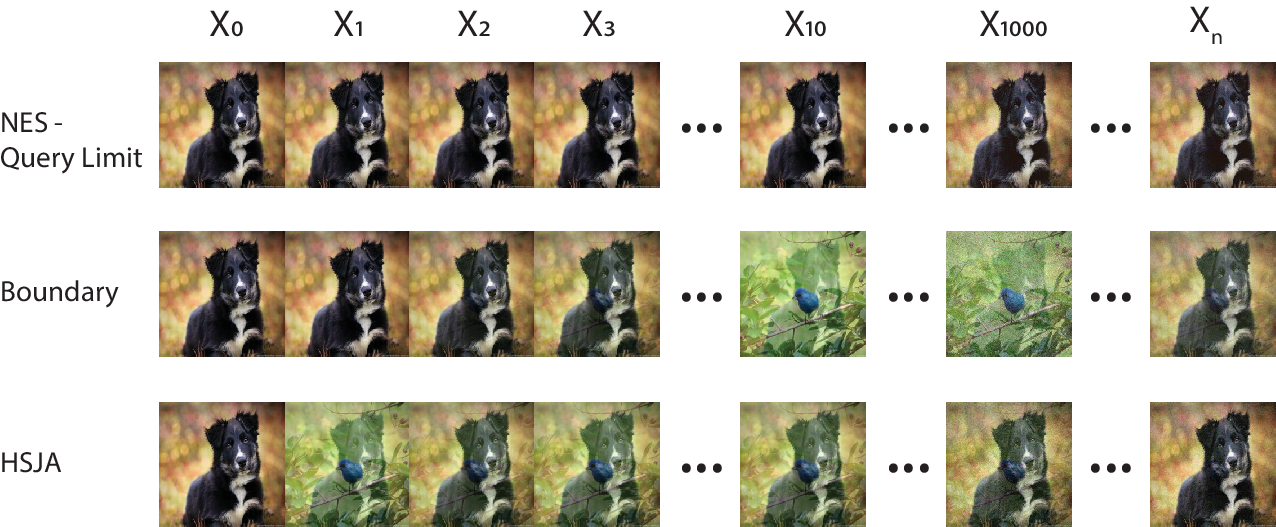}
   \vspace{-0.09in}
    \caption{\em Examples of attack query sequence $(x_0, x_1, ..,x_n)$,
      produced by three black-box attacks (NES, Boundary, HSJA).  While these
      attacks
      generate queries differently,  the resulting query sequences all contain
      some highly similar images.
    }
\vspace{-0.15in}
  \label{fig:attackimage}
\end{figure}

  We empirically verify the presence of highly similar queries by
  running the eight SOTA query-based black-box attacks (listed in
  Table~\ref{table:attack_category}) on
the \imagenet\ classification model. For all eight attacks, high
similarity is consistently observed across images in their attack
query sequence. The average $L_2$
distance between just consecutive queries in an attack sequence is
already 20-380x
smaller than analogous distance between benign images (estimated by
randomly comparing 2000 pairs of benign images).
Figure~\ref{fig:attackimage} shows some visual examples from attack query
sequences generated by three attacks (NES-Query Limit, Boundary,
HSJA). We omit the other attacks since they produce similar
results.








  \begin{figure*}[t]
  	\begin{minipage}[t]{0.75\linewidth}
      \raisebox{-\height+4\baselineskip}{\includegraphics[width=0.89\linewidth]{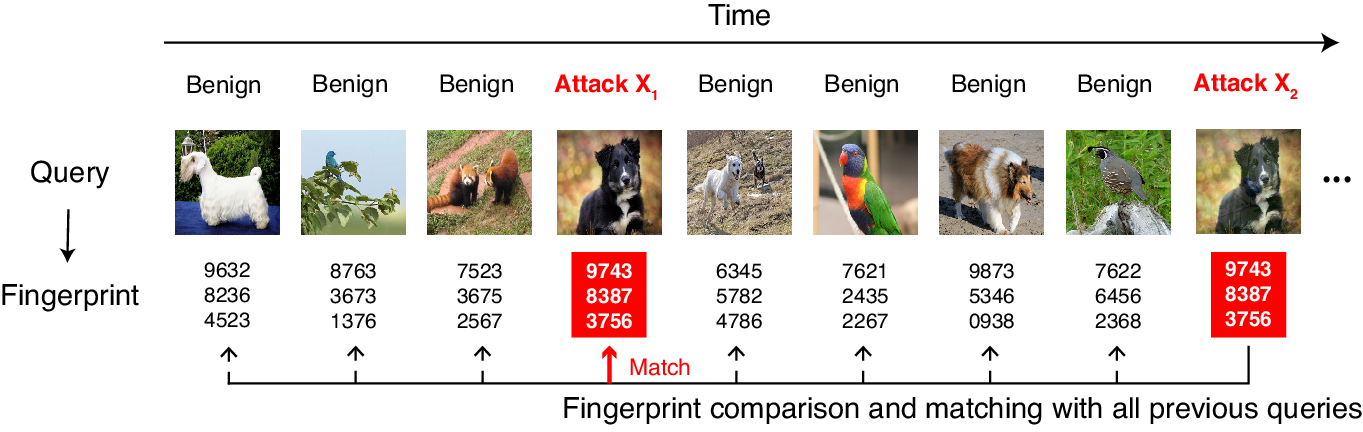}}
      \vspace{-0.1in}
      \captionof{figure}{\em For each raw image, Blacklight computes a small set of hash entries
        (as its probabilistic fingerprint). Blacklight detects attack images hidden
        inside a large stream of benign images by comparing and detecting highly
        similar fingerprints.}
      \label{fig:concept}
    \end{minipage}\hfill
  	\begin{minipage}[t]{0.23\linewidth}
    \centering
    \raisebox{-\height+3\baselineskip}{\includegraphics[width=.75\linewidth]{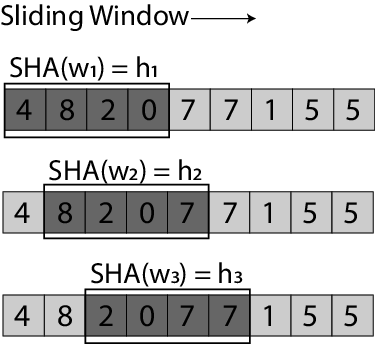}}
    \captionof{figure}{\em Computing content hashes by applying a sliding
      window over pixels.}
    \label{fig:detection_workflow}
    \end{minipage}
    \vspace{-0.15in}
  \end{figure*}

\vspace{-0.1in}
\subsection{Fast and Scalable Similarity Check via Probabilistic Fingerprinting}

The above insight motivates us to detect query-based black-box attacks by searching for the presence of highly similar queries in a large
stream of incoming \htedit{and past} queries. A key challenge is how to run a
fast and efficient similarity check.

\para{Strawman Solutions.} \htedit{We first discuss two strawman solutions and their
problems. Earlier in \S\ref{sec:existing} we discussed the
query similarity check used by SD~\cite{chen2020stateful} and its
scalability issue. }

\htedit{\spara{Computing $L_p$ distances.}} A naive approach would store all past queries in a database and
compare an incoming query $x$ to the entire database of $n$ queries by computing
their image-level differences. Such raw comparison incurs heavy costs both in query
storage and computation, \htedit{i.e., $O(n)$}.
For example,
even for low resolution image queries (224$\times$224 pixels, \imagenet), it takes
23 minutes to compare a query to one million prior images, even using five
threads on a 6-core Intel Xeon server.  This is clearly intractable in
practice.

\htedit{
\spara{Locality-sensitive (LS) hashing.}
An alternative is to compute a ``signature'' per query using
LS hashes and compare queries by their signatures.
Many have used perceptual
hashing (\eg PhotoDNA~\cite{photodna}, dhash~\cite{dhash}), a type of LS hashes,
to match similar images for copyright resolution or child exploitation
detection~\cite{photodna}. Using a hash table for lookup, the runtime cost for
checking each incoming query could reach $O(1)$ regardless of $n$.
Unfortunately, these hashes are designed to identify generic variants of
an image, even those that have
undergone significant alterations. Thus they flag similar benign queries (\eg different frames of a video, multiple pictures of
the same object) as adversarial, producing false positives. We test dhash~\cite{dhash} on our attack
detection
and find
that it produces over $10\%$ false positives on the Flickr dataset
and $67\%$ on our video dataset (\S\ref{sec:fpr}). While unable to
test PhotoDNA since it is proprietary, we expect that it faces the
same issue since it focuses on detecting child exploitation in images which requires considerable alterations.}

\para{Probabilistic Fingerprints.}  Blacklight overcomes these challenges by
applying probabilistic fingerprinting to detect highly similar
images. Our goal is to design a hash
function that is compact yet highly sensitive to very small changes in the
image. This dictates that we should use a highly lossy function. Probabilistic
fingerprinting achieves these properties and utilizes secure one-way hashes that
cannot be easily reversed to evade detection \htedit{and probabilistic
downsampling for efficiency}.  To fingerprint an image $x$, Blacklight first transforms $x$ into a set of
continuous and overlapping segments of a fixed length $\mathbf{w}$, then
applies a one-way hash to each segment to produce a large set of
$\mathbf{N}$ hash values.
From these $\mathbf{N}$ hash values, Blacklight chooses a small set
probabilistically (e.g., the top 50) as $x$'s probabilistic fingerprint.

Figure~\ref{fig:concept} illustrates Blacklight's attack detection process.
For an incoming query
$x$, Blacklight extracts its probabilistic fingerprint and stores it
in the database.  Blacklight runs an efficient hash match
algorithm to detect overlaps between $x$'s fingerprint and those in
the database.
Upon detecting sufficient overlap between $x$ and an existing
fingerprint $y$, it flags ($x$, $y$) as a pair of attack
queries.

\para{Key Benefits.}
Our  fingerprint scheme has the property that any two highly similar queries will produce a near-perfect match in their fingerprints. In other words,
small changes to an image are highly unlikely to impact its
fingerprint. The use of secure one-way hash and \htedit{probabilistic
  downsampling} means that unless they can reverse the hashing algorithm, an adversary
cannot alter an image's fingerprint without significantly
altering its content \htedit{(further confirmed in
\S\ref{sec:reduce_sim}).}


Our fingerprints also greatly reduce the storage
overhead of past queries, and the computation costs of comparing queries in
similarity. Specifically, the search for highly similar queries reduces down to a
hash set comparison problem, which takes near-constant
time in general (see \S\ref{sec:design}).



\para{Prior Work on  Probabilistic Fingerprints.} Probabilistic
fingerprints have been used for similarity detection in text (e.g., detecting code plagiarism~\cite{shivakumar1995scam, brin1995copy,
  roy2009comparison, ducasse1999language}, network intrusion and
malware~\cite{singh2004automated, roussev2009hashing, opricsa2016malware} and
spam emails~\cite{zhou2003approximate, liu2005detecting}). It was also
used in
\textit{sif}, a similarity detector for file
systems~\cite{manber1994finding}.  The contributions of our work
include i) extending probabilistic fingerprints beyond the text domain, ii) customizing its
design to identify similar image queries to a DNN model
(see \S\ref{sec:design}), and iii) a formal analysis to model both false
positives and attack detection rates and their dependency on
fingerprinting parameters (see \S\ref{sec:theory}).

\begin{figure*}[t]
  \centering
    \includegraphics[width=0.96\textwidth]{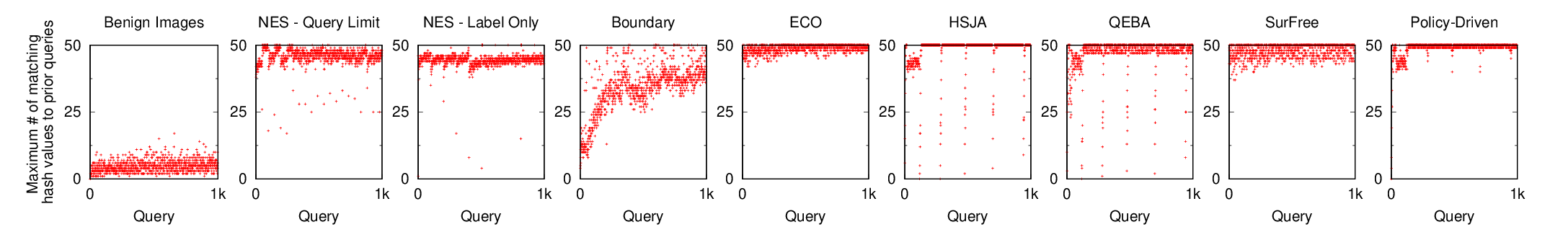}
    \vspace{-0.1in}
    \caption{\em We empirically show that probabilistic fingerprints
      preserve the query similarity in black-box attack
      sequences. We plot the {\em maximum fingerprint overlap} between $x$
      and that of any prior query in a benign query sequence (left most)
      and eight attack query sequences. Here the maximum matching is
      bounded by $\textbf{S}=50$. }
         \vspace{-0.2in}
  \label{fig:exampleDPF}
\end{figure*}

   \vspace{-0.12in}
\section{Detailed Design of Blacklight}
   \vspace{-0.1in}
\label{sec:design}

We now present the detailed design of Blacklight, including
preprocessing, probabilistic fingerprinting, and comparison algorithms, which
together form our proposed detector.  We also discuss options to mitigate
attacks after detection. Note that Blacklight works as an external
  add-on, and requires no modifications to the DNN model.

   \vspace{-0.1in}
   \subsection{Preprocessing: Salted Pixel Quantization}
      \vspace{-0.05in}
\label{subsec:quantization}

Given an incoming image query $x$, Blacklight first runs a quantization function on
each pixel of $x$.  This serves two purposes.  First, it converts continuous pixel values into a finite set of discrete values, which are then
used to compute hashes of $x$ during fingerprinting.  Second,
quantization increases similarity between (attack) queries. This is
particularly true for black-box attacks that iteratively optimize queries by
gradually modifying every single pixel on the image~\cite{ilyas2018black,
  chen2020hopskipjumpattack, brendel2018decision} -- the use of quantization
effectively nullifies changes to image hashes created by these minor
modifications without inducing false positives.
\htedit{We confirm this empirically in
  Figure~\ref{fig:quantization_match_comp} where the hash
  overlap between attack queries (on \cifar) increases
  rapidly with the quantization step $\mathbf{q}$ to approach 100\%, while those between benign
  queries remain low.}
Note that this step
  is used only for attack detection. If the input is
  considered benign, the original, unaltered query is sent to the DNN
  model.

Furthermore, Blacklight employs a {\em salted} pixel quantization
function to  resist reverse engineering attacks:
\begin{equation} \vspace{-0.1in}
  Q(x, salt_Q, \mathbf{q}) = \lfloor \frac{(x+salt_Q) \;\text{ mod }\; 255}{\mathbf{q}} \rfloor
  \label{eq:quantization}
\end{equation}
where $salt_Q$ is a randomly generated salt image (of the same dimensions
as $x$) and $\mathbf{q}$ is the quantization step (a system parameter).  Here
all pixel values of $x$ and $salt_Q$ are normalized to $[0,
255]$. Later in \S\ref{sec:countermeasure} we show adding a
  random salt improves Blacklight's robustness against adaptive attacks.


\vspace{-0.1in}
\subsection{Computing Probabilistic Fingerprints}
\label{subsec:fingerprint}
\vspace{-0.05in}
We now describe the detailed process to compute the probabilistic
fingerprint on a (quantized) query image $x$.

\para{Converting an image into $\textbf{N}$ segments.}  Blacklight first
``flattens'' the 2D image into a single pixel sequence by concatenating rows
of pixels together; then applies a sliding window of fixed size $\mathbf{w}$
on this sequence, iteratively moving the sliding window by $\mathbf{p}$
(referred to as the sliding step).  This produces $\mathbf{N}=(|x|-\mathbf{w}+\mathbf{p})/\mathbf{p}$
overlapping pixel segments, each of length $\mathbf{w}$.  Any two consecutive
segments overlap by $\mathbf{w-p}$ pixels, and each pixel in $x$ is included in
$\mathbf{w/p}$ segments.


\para{Hashing each segment.} For each segment $i$ ($i\in [1,\mathbf{N}]$),  Blacklight applies a secure
one-way hash function (e.g., SHA-3 combined with a random salt value chosen
by the defender) and produces a hash value $h_i$.
This creates a full hash set $\mathbf{H}_x=(h_1, h_2, ..., h_\mathbf{N})$ for query
$x$, with $\mathbf{N}$ hash entries. For example, for \cifar{} ($|x|=32 \times 32 \times 3= 3072$), $\mathbf{N}=3053$ when $\mathbf{w}=20$,
$\mathbf{p}=1$. An illustration of this sliding window hashing scheme is shown in
Figure~\ref{fig:detection_workflow}.

\para{Selecting a subset of hashes as the fingerprint.}
From $x$'s full hash set $\mathbf{H}_x$, Blacklight
selects the top $\mathbf{S}$ hash values (sorted by {\em numerical
  order}) as its probabilistic fingerprint, denoted as
$\mathbb{S}(\mathbf{H}_x)$.
Since the output distribution of the one-way hash is random, choosing the top
$\mathbf{S}$ hash values by numerical order serves as an efficient downsampling
algorithm that is \htedit{deterministic}\footnote{\htedit{
    Deterministic means that the downsampled hash set holds the same
    property of the full hash set: highly similar (quantized) queries
    will have highly similar fingerprints. We also verified this
    empirically in Figure~\ref{fig:l2_match}.}} to the defender but unpredictable to an adversary (since
predicting the top $\mathbf{S}$ hash values requires predicting the
full hash set).


The use of probabilistic fingerprinting puts a {\em hard} limit on the overhead of
fingerprint storage and comparison, while preserving the high similarity
among attack queries.  Figure~\ref{fig:exampleDPF} shows a sample
measurement on query similarity,  for the eight black-box attacks
discussed in \S\ref{sec:selected_attacks}.  Here we
measure, for each query $x_i$ in an attack sequence,  the maximum number of matching hashes between $x_i$'s fingerprint  and
any of its prior queries in the same sequence.  For reference,
we also compute the number of matching hashes among  benign images.
We see that many attack
queries display fingerprints highly similar to at least one
prior query in the same sequence, while benign queries
share minimal overlap in fingerprints.  Thus Blacklight can quickly detect black-box attacks after seeing only a small
number of queries.

\vspace{-0.15in}
\subsection{Comparing and Matching Fingerprints}
\vspace{-0.05in}
\label{subsec:comparing}

Upon receiving a new
query $x$, Blacklight computes its fingerprint $\mathbb{S}(\mathbf{H}_x)$
and compares it to all prior fingerprints stored in the database.  If any
stored fingerprint shares more than $\mathbf{T}$ hash entries with
$\mathbb{S}(\mathbf{H}_x)$, then $x$ is flagged as an attack image.  Here, the
value of $\mathbf{T}$ can be configured to meet the desired false positive rate. Later
in \S\ref{sec:theory}, we analytically show that by properly configuring $\mathbf{T}$
and $\mathbf{S}$, we can achieve accurate attack detection at a low false positive
rate.

Computing the maximum overlap between the fingerprint of a query and $n$
stored fingerprints
is non-trivial. A simple algorithm would
incur computation cost of $O(n)$. We use a better algorithm which
stores a query $x$'s fingerprint into a hashmap using each of its
  hash entry as a key.
The maximum overlap
with all $n$ queries can be found by retrieving all queries associated
with each key
 in $x$'s fingerprints, and counting the max frequency of any
query in that set. An efficient implementation can produce average
runtime that is a constant independent of $n$. We leave the design and
analysis of an efficient hashset matching algorithm to future work.
We present detailed performance overheads in \S\ref{sec:overhead}.

\vspace{-0.15in}
\subsection{Mitigating Attacks after Detection}
\vspace{-0.05in}
\label{sec:mitigation}

Detecting the presence of a query-based black-box attack is just a first step in
protecting DNN models. A persistent attacker can simply switch accounts and/or
IP addresses and continue with additional queries. Here, we
discuss options for mitigation after an attack is detected.

\para{Ban accounts.} As a response, banning an account or blocking an IP address
is not ideal. First, it means each false positive incurs a high penalty, which
might be undesirable in some application settings. Second, this does little
to deter resource rich attackers, who can continue the attack using Sybil
accounts, which are difficult to eradicate in practice.

\para{Return misguided outputs.} We also consider a more elaborate scheme where
the defender intentionally misleads the attacker by returning carefully
biased query outputs, perhaps towards secondary goals like identifying the
attacker. This approach faces additional challenges. First, crafting biased
responses requires significantly more computation and state-keeping at the
defender. Second, the defender must be careful to avoid returning valid
responses to actual attack queries.

\para{Reject all detected queries.} Ultimately we chose a simple strategy:
reject all detected attack queries. This mitigation is effective in
preventing attacks {\em IFF} the ratio of attack queries detected is high. If
most attack queries are rejected, the attack sequence takes a very long time
to converge and succeed. The benefit of this approach is that it does not
rely on detecting or reducing Sybil accounts, and false positives have minimal
impact on benign users.

\begin{figure}[t]
  \centering
    \includegraphics[width=0.4\textwidth]{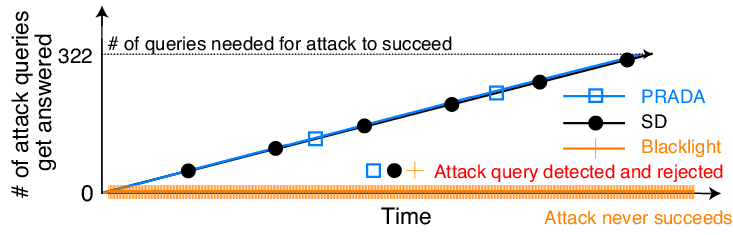}
    \vspace{-0.1in}
    \caption{\htedit{{\em By detecting/rejecting most of attack queries (regardless
      of account usage), Blacklight effectively resists persist
      attackers, which existing defenses fail to address.}}}
         \vspace{-0.2in}
  \label{fig:compare_trace}
\end{figure}

\htedit{ In \S\ref{sec:eval}, we evaluate the impact of mitigation on
persistent attackers who continue
to submit attack queries after query rejection.
Figure~\ref{fig:compare_trace} provides a preview in terms of the \#
of attack
queries got answered under Blacklight, using the
persistent attack trace of Figure~\ref{fig:sdtrace}. Blacklight rejects almost all the attack queries,
preventing the attack from making progress. }



 \vspace{-0.15in}
\section{Formal Analysis}
\vspace{-0.1in}
\label{sec:theory}
We formally examine Blacklight by
modeling the process of probabilistic fingerprinting. We derive analytical bounds on the probability of Blacklight
flagging a query pair $(x,y)$ as attacks $Q(\Delta)$ as a function of the full
hash difference between the two, i.e.,
$\Delta=\text{diff}(\mathbf{H}_x,\mathbf{H}_y)$.  We then estimate
                        Blacklight's false positive rate and attack
                        detection rate by  $Q(\Delta_{benign})$ and
                        $Q(\Delta_{attack})$. Here $\Delta_{benign}$ is the minimum full hash difference between
benign queries and $\Delta_{attack}$ is the maximum full hash difference between
attack queries. Our key results:  are:  (i) $Q(\Delta)$ decays fast with
$\Delta$, (ii) Blacklight can effectively
detect attacks at a low false
positive rate:  $Q(\Delta_{attack})\rightarrow 1,
\;Q(\Delta_{benign})\rightarrow 0$,  if $\Delta_{benign}
>>\Delta_{attack}$, (iii) \htedit{the analytical bound on  $Q(\Delta)$ can
guide the selection of Blacklight's configuration parameters
($\mathbf{w}$, $\mathbf{p}$, $\mathbf{q}$,  $\mathbf{S}$ and
$\mathbf{T}$). For brevity, we leave the details to
Appendix\S\ref{sec:theory_app}. }

\vspace{-0.1in}
\section{Experimental Evaluation}
\label{sec:eval}
\vspace{-0.05in}
Using four different image classification tasks (and datasets),  we
empirically evaluate Blacklight against  eight SOTA black-box
attacks. Our experiments seek to
understand 1) the effectiveness of Blacklight in both attack
detection and mitigation; 2) the false positive rate under realistic settings;
3) impact of Blacklight configuration; 4) Blacklight's storage and computation cost; 5) applying Blacklight to other
domain.

\vspace{-0.1in}
\subsection{Experimental Setup}
\vspace{-0.05in}
\label{sec:tasks}
We apply Blacklight to protect DNN models
developed for image classification. Our experiments cover a wide range of
input size/content and model architectures, allowing us to evaluate
Blacklight under a diverse set of conditions.


\para{Image Classification Tasks.} We consider four representative tasks: \mnist~\cite{lecun1998gradient},
\gtsrb~\cite{Stallkamp2012}, \cifar~\cite{krizhevsky2009learning} and
\imagenet~\cite{ILSVRC15}. We summarize in Appendix \S\ref{sec:extra_config} these tasks and associated models in Table~\ref{table:tasks}, and detailed model architectures and training
configurations in Table~\ref{table:model_mnist} to~\ref{tab:train_detail}.

\begin{figure*}[t]
	\begin{minipage}[t]{0.6\linewidth}
    \centering
      \resizebox{\textwidth}{!}{
          \begin{tabular}{@{\extracolsep{1pt}}l|l|ccc|c|cc}
              \hline
              \multirow{2}{*}{Task} &
              \multirow{2}{*}{Attack} & \multicolumn{3}{c|}{w. Detection} &    \multicolumn{1}{c|}{w. Mitigation}                                                                                                                                                                                                              & \multicolumn{2}{c}{w/o Blacklight}                                                                                       \\\cline{3-8}
                                      & & \begin{tabular}[c]{@{}l@{}}Attack\\detect \%\end{tabular} & \begin{tabular}[c]{@{}l@{}}Detection\\coverage\end{tabular} & \begin{tabular}[c]{@{}l@{}}Avg queries\\to detection\end{tabular} & \begin{tabular}[c]{@{}l@{}}Attack\\success\end{tabular} & \begin{tabular}[c]{@{}l@{}}Attack\\success\end{tabular} & \begin{tabular}[c]{@{}l@{}}Avg \# attack\\ queries\end{tabular}  \\
                         \hline
\multirow{8}{*}{\mnist}
              &NES - QL & 100\% & 99.5\% & 2 & 0\% & 45\%  & 66540\\
              &NES - LO  & 100\% & 99.0\% & 2 & 0\% & 1\%  & 95973\\
              &Boundary          & 100\% & 64.2\% & 18 & 0\% &  21\% & 85467\\
              &ECO               & 100\% & 99.9\% & 2 & 0\% & 43\%  & 52780\\
              &HSJA              & 100\% & 98.1\% & 6 & 0\% & 59\%  & 9924\\
              &QEBA              & 100\% & 98.4\% & 8 & 0\% & 92\%  & 12141\\
              &SurFree              & 100\% & 97.9\% & 7 & 0\% & 84\%  & 10034\\
              &Policy-Driven              & 100\% & 99.0\% & 8 & 0\% & 74\%  & 9538\\
              \hline

\multirow{8}{*}{\gtsrb}
            &NES - QL & 100\% & 98.5\% & 2 & 0\% & 66\%  & 48429\\
            &NES - LO  & 100\% & 98.0\% & 3 & 0\% & 17\%  & 83823\\
            &Boundary          & 100\% & 64.3\% & 22 & 0\% & 37\% & 76643\\
            &ECO               & 100\% & 100.0\% & 2 & 0\% & 80\%  & 27782\\
            &HSJA              & 100\% & 98.2\% & 5 & 0\% & 95\%  & 10392\\
            &QEBA              & 100\% & 99.5\% & 8 & 0\% & 99\%  & 9832\\
            &SurFree           & 100\% & 98.3\% & 8 & 0\% & 98\%  & 9192\\
            &Policy-Driven     & 100\% & 98.1\% & 5 & 0\% & 100\%  & 13021\\
            \hline

\multirow{8}{*}{\cifar}
          &NES - QL & 100\% & 98.3\% & 2 & 0\% & 100\%  & 12621 \\
          &NES - LO  & 100\% & 98.7\% & 2 & 0\% & 89\% & 67126 \\
          &Boundary          & 100\% & 64.4\% & 25 & 0\% & 95\% & 6082 \\
          &ECO               & 100\% & 99.4\% & 2 & 0\% & 89\% & 16887 \\
          &HSJA              & 100\% & 97.1\% & 7 & 0\% & 100\% & 1205 \\
          &QEBA              & 100\% & 96.9\% & 6 & 0\% & 99\%  & 1009 \\
          &SurFree           & 100\% & 96.8\% & 8 & 0\% & 100\%  & 1396 \\
          &Policy-Driven     & 100\% & 97.3\% & 7 & 0\% & 100\%  & 1198 \\
          \hline

\multirow{8}{*}{\imagenet}
          &NES - QL & 100\% & 99.4\% & 2 & 0\% & 99\%  & 11201 \\
          &NES - LO  & 100\% & 98.2\% & 2 & 0\% & 20\%  & 63492 \\
          &Boundary          & 100\% & 95.1\% & 42 & 0\% & 74\%  & 67356 \\
          &ECO               & 100\% & 99.6\% & 2 & 0\% & 93\%  & 11304 \\
          &HSJA              & 100\% & 98.7\% & 7 & 0\% & 99\%  & 12402 \\
          &QEBA              & 100\% & 98.3\% & 6 & 0\% & 100\%  & 10293\\
          &SurFree           & 100\% & 97.6\% & 7 & 0\% & 100\%  & 8783\\
          &Policy-Driven     & 100\% & 99.1\% & 8 & 0\% & 100\%  & 10368\\
         \hline
      \end{tabular}}
\vspace{-0.1in}
  \captionof{table}{\em Blacklight's detection and mitigation
    results. In the last two columns, we included attack performance
   in absence of Blacklight: attack success rate
    and average attack queries required to complete an attack.}
    \label{table:results}
  \end{minipage}
  \hfill
    \begin{minipage}[t]{0.38\linewidth}
      \vspace{-145pt}
      \centering
      \raisebox{-\height+4\baselineskip}{\includegraphics[width=0.9\linewidth]{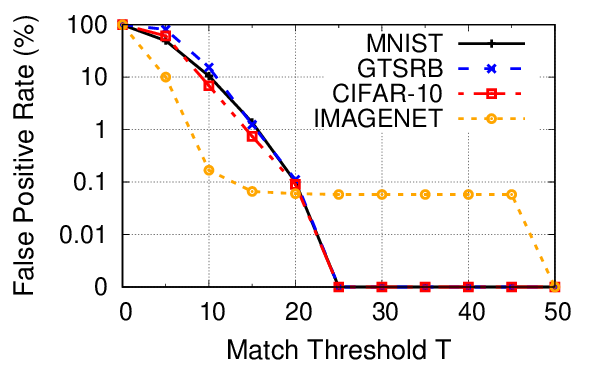}}
      \vspace{-0.1in}
      \captionof{figure}{\em Blacklight's false positive rate
       when fixing $\mathbf{S}=50$ and varying $\mathbf{T}$.}
      \label{fig:false_positive}

    \vfill
\vspace{8pt}
    \centering
    \resizebox{\textwidth}{!}{
      \begin{tabular}{l|l|l||l|l|l}
      \hline
      Label     & \# of Filtered & FPR    & Label     & \# of Filtered & FPR    \\ \hline
      balloon   & 953            & 0.07\% & packet    & 1158           & 0.21\% \\
      boathouse & 1572           & 0.73\% & peacock   & 556            & 0.68\% \\
      daisy     & 656            & 0.10\% & pier      & 309            & 0.07\% \\
      fly       & 188            & 0.03\% & rifle     & 905            & 0.48\% \\
      geyser    & 896            & 0.11\% & snail     & 350            & 1.01\% \\
      hay       & 1192           & 0.79\% & swing     & 510            & 0.48\% \\
      knot      & 817            & 0.14\% & teapot    & 1715           & 0.14\% \\
      menu      & 1232           & 0.37\% & tiger cat & 1315           & 0.28\% \\
      mortar    & 1229           & 0.38\% & toaster   & 3298           & 0.37\% \\
      nail      & 1696           & 0.83\% & vault     & 182            & 0.04\% \\ \hline
      \end{tabular}}
    \captionof{table}{\em Blacklight's false positives on
      benign images crawled from Flickr. ``\# of Filtered''
    is \# of images that are duplicated and have the same hash
    value with prior queries;
    ``FPR'' is the false positive rate per label.  For each label, we
    run Blacklight on
    80,000 Flickr images (crawled via this label).} 
    \label{table:flickr_fpr}
  \end{minipage}
  \vspace{-0.1in}
\end{figure*}

\para{Attack Configurations.} We implement and run the eight
black-box attacks list in Table~\ref{table:attack_category} against each of the above four classification
models.   For  \mnist, \gtsrb~and \cifar, we randomly select 1000
images from their test datasets and use each
as the source image of the attack ({\em i.e.\/} $x_0$). For \imagenet,
we randomly select
500 source images (due to its higher computation cost).
We run each attack until it
terminates (i.e., successfully generating an
adversarial example) or reaches 100K queries, whichever occurs first.

When configuring each attack, we follow its original paper and
  use the same L$_p$ distance metric (L$_2$ or L$_\infty$) stated in
  the paper.
Since L$_2$ distance depends on model input size,
we use the {\em normalized} L$_2$ distance \htedit{$\sqrt{\frac{1}{|x|}\sum_{i=0}^{|x|}(x_i - x_i')^2}$.}
  The detailed attack parameters and distance metrics are listed in
  Table~\ref{table:attackconfig}.

For all these attacks, we set the perturbation budget
$\epsilon$ such that most attacks
succeed in absence of defenses.  As reference, the standard $\epsilon$
for white-box attacks is 0.03  for L$_\infty$ and  $<$0.03 for {\em normalized}
L$_2$~\cite{carlini2017towards}. Black-box attacks should
use a larger budget because they are naturally harder to succeed. In
fact, our experiments on the eight SOTA black-box attacks
confirm that a budget of 0.03 leads to significant attack
failures. Thus we increase $\epsilon$=0.05 for
both L$_\infty$ and {\em normalized} L$_2$ to allow most attacks to succeed.  The
only exceptions are  L$_\infty$ attacks against \mnist\ since
$\epsilon$=0.1 is necessary for them to succeed.  The perturbation budgets are listed in
Table~\ref{table:attackconfig}.

\para{Blacklight Configuration.}  Table~\ref{table:configuration} in Appendix
  lists the default values for Blacklight's key parameters: sliding
  window size ($\mathbf{w}$), sliding step ($\mathbf{p}$), quantization step ($\mathbf{q}$), \# of hash
  entries per fingerprint ($\mathbf{S}$), and fingerprint matching
  threshold ($\mathbf{T}$).  To demonstrate the generality of Blacklight, we set these parameters to be the same
  default values for all four tasks, rather than ``optimizing'' them
  per
  task.  The only exception is $\mathbf{w}$  -- our default
  value is $20$, but we increase it to 50 for \mnist{} (due to its large
  black background) and \imagenet{} (due to its large image size).

\htedit{We choose these values following our formal analysis. In particular, we
 choose $\mathbf{T}=\mathbf{S}/2=25$ by modeling how $\mathbf{T}$
affects false positive and detection coverage.  Figure~\ref{fig:false_positive} shows the measured
false positive rates when varying
$\mathbf{T}$, confirming that $\mathbf{T}$=25 achieves less than 0.1\%
false
positive for all four tasks.  In \S\ref{sec:param_analysis}, we
further explore the impact of
  parameter configuration by varying
  $\mathbf{w}$,  $\mathbf{p}$,  $\mathbf{q}$ and $\mathbf{S}$.}

\para{Evaluation Metrics.}
We use the following metrics to quantify the
effectiveness and cost of Blacklight.
\begin{packed_itemize}\vspace{-0.1in}
  \item {\bf False positive rate}: \% of benign queries detected as
    attack.
\item {\bf Attack detection rate:} \% of black-box attacks
  detected before the attack completes.
\item {\bf Detection coverage:} \% of queries in an
  attack's query sequence identified as attack queries. 
\item {\bf Avg \# of queries to detection:} \htedit{Average \# of attack queries
  accepted (thus answered) before detecting an attack query}.
\item {\bf Attack success rate w. mitigation:} Success rate of a persistent
  attack when all detected attack queries are rejected.
\item {\bf Detection overhead:} Run-time latency and storage costs. \vspace{-0.05in}
\end{packed_itemize}

\vspace{-0.15in}
\subsection{Attack Detection and Mitigation}
\vspace{-0.05in}
\label{subsec:8.2}
We evaluate Blacklight's detection rate by implementing and performing each
of the eight black-box attacks against each classification model. For each
attack and task combination, we run 1000 instances of the attack (500 for
\imagenet). Each attack instance selects a random image
from the test dataset as source image of the attack ($x_0$), and a random
incorrect label as the misclassification target label.

The results for all attacks are listed in
Table~\ref{table:results}.  As reference, the last two columns report the performance
of these attacks {\em without} the Blacklight defense, in
terms of attack success rate and the speed of convergence (\# of
queries before successfully producing an adversarial example).  We
see that recent attacks, especially HSJA, QEBA, SurFree, Policy-Driven,
are highly successful in absence of Blacklight. Boundary and NES-LO take the longest
  time to converge. Some attack instances
do fail to converge even after generating 100k
queries (e.g.,  less than 50\% of NES-LO complete in 100K queries for
  MNIST, GTSRB and ImageNet).  Most of them remain unsuccessful even
  when increasing the query bound to 300k.  Overall,
a successful attack takes several thousands to tens of thousands of
queries to complete.

Next, we summarize key results on Blacklight's attack detection (as
shown by  column 3-5 in Table~\ref{table:results}).  We see that the attack
detection rate remains 100\% for all attack instances, indicating that
Blacklight detects {\em all} attacks on all models in progress.
The detection coverage is also extremely high -- Blacklight detects more than 96\% of all attack queries, except on the
Boundary attack.  Another key observation is that
Blacklight detects a new attack instance very
quickly, often after a handful of 2--8 queries (again, more queries
required for Boundary because it converges slower).  In all cases, Blacklight
detects an attack in less than 1\% of the average number of queries
required to complete the attack.

\htedit{Blacklight detects Boundary slower than others. This is
  because  Boundary advances slower in shrinking perturbation towards the $L_p$ ball of the
  target, thus Blacklight detects them at a ``later'' stage with 100\% detection rate.  The three improved versions of Boundary
  (HSJA, QEBA, Policy) converge faster, thus Blacklight detects
  them faster.  To further evaluate the slower Boundary attack, we run the
  attack for 1 million queries. We find Blacklight continues to
  detect (and reject) attack queries in this longer sequence, leaving
  the
  attacker with 0\% success (for all four tasks). The
  detailed results are listed in Table~\ref{table:boundary_results} in Appendix.}


Finally, column 6 in Table~\ref{table:results} reports the attack
success rate when Blacklight rejects queries identified as
attack queries. We see that none (0\%) of
persistent attackers manage to complete their attack within 100K
queries.  Blacklight's mitigation is highly effective because it is able to
detect nearly all attack queries. Rejecting these queries
prevents the attacker from making forward progress in probing
model classification boundaries.  This confirms that a high detection coverage
is critical to defend against query-based black-box attacks.

\para{Key Takeaways.} Our results against eight SOTA black-box attacks
show that Blacklight detects all attacks on all
models, detects the overwhelming majority of queries in the attack
sequence, and detects the attack quickly (usually in less than 8 queries,
with the exception of the slow converging Boundary attack).
Furthermore, by rejecting all detected attack queries, Blacklight's
mitigation module ensures no attacks can complete (at least in 100K
queries)  for all our tested attacks and target DNN models.

\htedit{\para{Comparison to Existing Defenses.} As reference, we show
  the performance of SD and
  PRADA in Table~\ref{table:compare}, using the same attack
  experiments described above. As discussed in
  \S\ref{sec:existing}, SD and PRADA are not designed
  to stop  persistent attackers who switch account to continue attack.   Results in
  Table~\ref{table:compare} confirm this and their low
  detection coverage (0.8\%-2.1\%).
  }

\vspace{-0.15in}
\subsection{Detecting Universal Patch Attacks}
\label{sec:uni_patch}
\vspace{-0.05in}
\label{subsec:patchresult} We evaluate Blacklight
  against the only known query-based universal patch attack,
  Sparse-RS~\cite{croce2020sparse}. Table~\ref{tab:sparse-rs} in Appendix
  shows that Blacklight is also highly effective in detecting
  Sparse-RS (100\% detection success rate and $>$ 97.6\% detection
  coverage).  Since query-based universal patch attacks are emerging,
  additional work is required to thoroughly evaluate the robustness of Blacklight against them.

\vspace{-0.15in}
\subsection{False Positives in Real World Settings}
\vspace{-0.05in}
\label{sec:fpr}

Since Blacklight relies on a similarity detection algorithm to detect
attacks, one might wonder if duplicates or near-duplicates of images will
trigger false positives.  Figure~\ref{fig:false_positive} reports
its false positives between distinctive inputs. But what about
``naturally'' similar images, such as different versions of the same image,
or closeby frames of the same video?

We begin with a simple test to confirm that naturally occurring false
positives are very low in large image repositories like ImageNet. We turn
off \htedit{database} resets, randomly sample 1 million images from
ImageNet training data, send them as queries to Blacklight, and observe a
very low false positive rate of $0.37\%$.

\para{False Positives in Similar Images.} Next, we look at similar images of
the same objects, e.g. inputs that should classify to the same labels.
We crawl a large number of public real world images from
Flickr~\cite{flickr_2020} using keyword search on their public API. We pick
20 random labels from ImageNet, and use each as a search keyword to crawl
$80,000$ images for that label.  We filter out images that are perfectly
identical at the pixel level (we found an average of $1036 \pm 696$ duplicate
images per label). We then take each label, and run our $80,000$ images as
queries to Blacklight. Even across Flickr images labeled with the same
keyword, Blacklight produces a very low false positive rate of $0.37\% \pm 0.29$
over 20 labels.  Detailed results for all labels are shown in
Table~\ref{table:flickr_fpr}.

\para{False Positives in Video Frames.} Finally, we consider the scenario
where the system might receive benign queries that are highly similar by
nature, e.g. image stills taken from video frames. We explore how Blacklight
responds under such scenarios by testing it for false positives on the
YouTube Faces dataset~\cite{wolf2011face}. YouTube Faces is a collection of
$3,425$ videos of $1,595$ different people, designed for studying
unconstrained facial recognition. We use common image extraction
techniques~\cite{wang2019neural} to extract $587,137$ video frame images from
videos for $1,283$ celebrities. Of these, we filter out $33,227$ images that
are pixel-level identical to other images, and send the remaining video
frames to Blacklight. The result is a false positive rate of $1.74\%$. Even if Blacklight
takes over half million queries per reset cycle for the highly similar
queries, the false positive rate is still very low.

\htedit{
\vspace{-0.15in}
\subsection{Impact of Parameter Configuration}
\label{sec:param_analysis}
\vspace{-0.06in}
As discussed in Appendix\ref{sec:theory_app}, we leverage our formal
analysis of Blacklight to configure its five system parameters:
$\mathbf{w}$, $\mathbf{p}$, $\mathbf{q}$, $\mathbf{S}$, and
$\mathbf{T}$.  Earlier in Figure~\ref{fig:false_positive} we show
empirically how
Blacklight's false positive rate varies with $\mathbf{T}$ and
verify our strategy on configuring $\mathbf{T}$. In the following, we study the impact of the other four parameters by testing
Blacklight against the same set of attacks while varying each
of these parameters.  We report the false positive rate and detection coverage
since the attack detection rate is always 100\%.  The detailed results
are listed in Figure~\ref{fig:parameter_analysis}  in Appendix.

We
summarize the key findings below.  First, we confirm that $\mathbf{q}$ is a critical parameter for Blacklight --
the
detection coverage increases quickly as $\mathbf{q}$ goes from 1 (no
quantization) to
50 (the default value) and stabilizes after that (except for
Boundary). When $\mathbf{q}$ approaches 100, we start to see visible increase in
false positives ($>$0.1\%). Second, as expected, the sliding window size
$\mathbf{w}$ is negatively correlated to false positive rate and detection
coverage, while the sliding step $\mathbf{p}$ has little impact
(note that $\mathbf{p}<\mathbf{w}$).
Thus Blacklight should select
$\mathbf{w}$ as a small value to meet the desired false
positive rate.
Finally, as expected $\mathbf{S}$ should be small
to reduce complexity but not too small (e.g., $<$20) to introduce visible
false positives. Overall, these results confirm our proposed
theory-guided
principle for choosing Blacklight's parameters.

}

\vspace{-0.1in}
\subsection{Overhead of Blacklight}
\label{sec:overhead}
\vspace{-0.08in}
\htedit{
\para{Storage.} Blacklight requires a database to store
fingerprints of prior queries. Our probabilistic fingerprints are
extremely small. Across all of our experiments, a fingerprint
is $\leq 32 \cdot \mathbf{S}$ bytes and 1.6KB for the default
configuration in Table~\ref{table:configuration}.
A database of $1$ million queries only requires 2GB
storage, a ``negligible'' value for modern servers.
}

\para{Runtime.}  Blacklight's per-query runtime includes
latency to generate the fingerprint from a query and latency to lookup the
fingerprint in the query database. The former depends on the image
size and the parameters ($\mathbf{w}$,
$\mathbf{p}$) and the latter depends on the size of query database
$n$.  We configure Blacklight to its default
configuration and explore the
impact of sliding step $\mathbf{p}$ (i.e.,
increasing $\mathbf{p}$ from 1 to 10 or 25 to speed up
hash computation) and the query database size $n$.
\htedit{We run Blacklight on an Intel i7
desktop server with $64$ GB memory, and report the per-query runtime
for two types of query images ($32\times 32$,\cifar{}) and ($224
\times 224$, \imagenet{}) in
Figure~\ref{fig:overhead_time} as a function of $n$. The curves remain flat over
$n$, suggesting that Blacklight's detection cost is independent of
$n$.  More specifically, a \cifar{} model inference takes 50ms (on a
Nvidia Titan RTX) while Blacklight (on Intel i7) takes 4-8ms
(8\%-16\% over 50ms) for
$n$=1 million queries.

\begin{figure}[t]
    \centering
    {\includegraphics[width=0.95\linewidth]{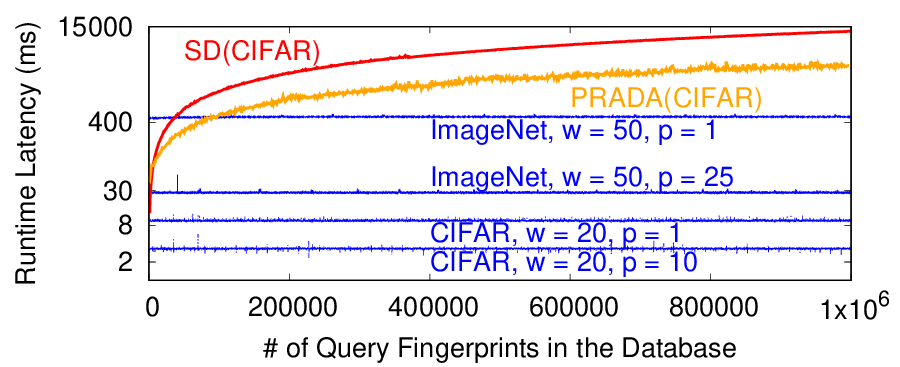}}
     \vspace{-0.1in}
     \caption{\em Blacklight's runtime latency vs. $n$. Note the log Y
      axis. We include \htedit{latency of SD and PRADA for reference}.}
      \label{fig:overhead_time}   \vspace{-0.2in}
\end{figure}

As reference, we compute the runtime
of SD and PRADA on the same Intel i7 server, putting all $n$ queries into a single
account.  They only run on \cifar{}, which we  report in
Figure~\ref{fig:overhead_time}.  The latencies scale linearly with $n$ (note the log Y
axis).  For $n$=1 million queries, SD and PRADA take 12s and 3.4s per
query (24,000\% and 6,800\% over inference).

\para{Further optimization.}  Blacklight's per query
latency is dominated by the sliding window-based hash computation
(99\% of total runtime).  We further optimize this computation using
GPUs. A modified version of Blacklight running a Nvidia
Titan RTX reduces per-query latency by 20x, to 0.4ms
for \cifar{} and 20ms for \imagenet{},  almost ``negligible''
compared to the inference latency.

}

\htedit{
\vspace{-0.15in}
\subsection{Blacklight for Text Classification}
\label{sec:other_domain}
\vspace{-0.05in}
Blacklight should in principle extend to other
domains where black-box adversarial attacks produce highly similar
queries in the input space. The domain-specific design task is how to generate query fingerprints to enable efficient and accurate
detection. Below, we show an initial Blacklight design for text
classification, a critical task in NLP.  DNN-based text classification is
shown to be vulnerable to query-based black-box
attacks~\cite{li2019textbugger, jin2020bert, maheshwary2021generating,
  gao2018black}, with three SOTA attacks: TextFooler~\cite{jin2020bert}, TextBugger~\cite{li2019textbugger} and
HardLabel~\cite{maheshwary2021generating}.

\para{Fingerprinting a sentence.}  The input to a text classifier is a
sentence, from which Blacklight
produces a fingerprint. First, we convert the
sentence into an array by replacing each word with its word
embedding. We quantize the array, apply a sliding window to move
through the quantized array and compute hashes, and select the top
$\mathbf{S}$ hashes as the query
fingerprint.  The parameter choices are listed in
Table~\ref{table:configuration} for \imdb{} text queries. $\mathbf{S}$
and $w$ are smaller
since text sentences create ``shorter'' arrays, while
$\mathbf{T}$ remains $\mathbf{S}/2$.

\para{Blacklight performance.} We run Blacklight on the three SOTA
attacks on the \imdb{}
dataset~\cite{maas2011learning}.  The results
in Table~\ref{table:nlp} show that Blacklight achieves 100\% detection
rate and $>$99.7\% detection coverage,  only takes 2 queries to
detect an attack (and reject the second query).  As such, no attack ever succeeds. For all of these tests, the false positive rate is only
0.49\%.  Overall, these results offer clear evidence that Blacklight
can potentially generalize to other domains using the same
probabilistic fingerprint methodology.
}

\begin{table}[t]
  \centering
  \resizebox{.48\textwidth}{!}{
  \begin{tabular}{c|ccc|c|cc}
      \hline
  \multirow{2}{*}{Attack} & \multicolumn{3}{c|}{w. Detection} &    \multicolumn{1}{c|}{w. Mitigation}                                                                                                                                                                                                               & \multicolumn{2}{c}{w/o Blacklight}                                                                                       \\\cline{2-7}
                          & \begin{tabular}[c]{@{}l@{}}Attack\\detect \%\end{tabular} & \begin{tabular}[c]{@{}l@{}}Detection\\coverage\end{tabular} & \begin{tabular}[c]{@{}l@{}}Avg queries\\to detection\end{tabular} & \begin{tabular}[c]{@{}l@{}}Attack\\success\end{tabular} & \begin{tabular}[c]{@{}l@{}}Attack\\success\end{tabular} & \begin{tabular}[c]{@{}l@{}}Avg \# attack\\ queries\end{tabular}  \\
             \hline

               TextBugger~\cite{li2019textbugger} & 100\% & 99.7\% & 2 & 0\% & 86.0\%  & 537\\
               TextFooler~\cite{jin2020bert}  & 100\% & 99.7\% & 2 & 0\% & 100.0\%  & 669\\
               Hard Label~\cite{maheshwary2021generating} & 100\% & 99.9\% & 2 & 0\% & 100.0\%  & 4642\\
\hline
\end{tabular}
}
\vspace{-0.1in}
\caption{\htedit{\em Blacklight’s detection and mitigation results on
    query-based
    black-box attacks for text classification.} }
\label{table:nlp}
\vspace{-0.2in}
\end{table}

\vspace{-0.15in}
\section{Adaptive Attacks}
\vspace{-0.1in}
\label{sec:countermeasure}
A meaningful defense must be robust against adaptive countermeasures from
attackers with full knowledge of the defense. We explored a number of customized
adaptive attacks against Blacklight, and present the strongest
countermeasures, organized into three groups: 1)
reducing query similarity for attack sequences, 2) reducing queries
needed for successful attacks and 3) leveraging resets in
Blacklight. Given the similarity between the attacks, we only apply
countermeasures to 5 of 8 attacks: NES (QL \& LO),
Boundary, ECO and HSJA.

\vspace{-0.1in}
\subsection{Reducing Query Similarity}
\vspace{-0.05in}
\label{sec:reduce_sim}
With knowledge of how Blacklight works, the straightforward adaptive attack
is to evade detection by reducing similarity between attack queries.
Below we present four types of adaptive attacks that add perturbations to attack
queries to reduce similarity between them.

\para{Evasion via Image Transformations.}  An attacker can try to evade
detection by adding additional perturbations to attack queries, where
ideally these perturbations do not disrupt the iterative optimization process, but are significant
enough to make fingerprints of attack queries different.  We explore
two types of image transformations: 1) adding Gaussian noise, and 2) applying
image augmentation like shift, rotation, zoom and
blending. \htedit{We apply these transformations to attack queries and
  send them to Blacklight. We first examine how these transformations
  affect the attack in absence of Blacklight, and confirm that they do
  introduce different levels
  of disruptions (none to 100\%).  On the other hand, for all the
  transformed attack sequences that will lead to a
  successful attack in absence of Blacklight,  Blacklight
  detects all of them, i.e., 100\% attack detection rate.
  Further details are in Appendix\S~\ref{sec:additional_results_image_trans} and
  Table~\ref{tab:transformation}.
}


\para{Increasing Learning Rates.}
The attacker can also try to increase dissimilarity between consecutive queries
 by tweaking their learning rate parameter.  Learning rate controls the
difference between two adjacent queries when estimating gradients.  This
does not apply to gradient estimation free attacks (Boundary and ECO). We
only explore different learning rate for
NES-QL, NES-LO and HSJA attacks.  For two variants of NES,
we gradually increase learning rate more than
$1000$ fold.  While the attack success rate drops to $0\%$, detection success
rate remains $100\%$.  For HSJA, we gradually grow learning rate
up to a factor of $10^6$, until changes in learning rate no longer impact
gradient estimation results. Here, attack success rate
steadily drops (eventually to 15\%), but detection remains at 100\%
throughout.

\begin{table}[t]
  \centering
  \resizebox{.45\textwidth}{!}{
  \begin{tabular}{l|cccc|cccc}
  \hline
  \multirow{2}{*}{\begin{tabular}[c]{@{}l@{}}Attack \\ Type\end{tabular}} & \multicolumn{4}{c|}{Default $\mathbf{T}=25$ (FPR = 0.0\%)} & \multicolumn{4}{c}{$\mathbf{T}=15$ (FPR = 0.74\%)} \\ \cline{2-9}
                                                                          & 0.05        & 0.1         & 0.15       & 0.2        & 0.05        & 0.1        & 0.15       & 0.2        \\ \hline
  NES - QL                                                                & 100\%       & 100\%       & 100\%      & 100\%      & 100\%       & 100\%      & 100\%      & 100\%      \\ \hline
  NES - LO                                                                & 100\%       & 100\%       & 100\%      & 100\%      & 100\%       & 100\%      & 100\%      & 100\%      \\ \hline
  Boundary                                                                & 100\%       & 100\%       & 75\%       & 40\%       & 100\%       & 100\%      & 100\%      & 95\%       \\ \hline
  ECO                                                                     & 100\%       & 100\%       & 100\%      & 100\%      & 100\%       & 100\%      & 100\%      & 100\%      \\ \hline
  HSJA                                                                    & 100\%       & 100\%       & 55\%       & 40\%       & 100\%       & 100\%      & 80\%       & 40\%       \\ \hline
  \end{tabular}}
\vspace{-0.1in}
\caption{\em Blacklight detection rate for attacks using larger perturbation
  budgets (0.1-0.2) for \cifar{}. Lowering $\mathbf{T}$ largely improves detection
  when attackers operate on very large perturbations, with a small increase in false
  positives.}
\label{table:perturb}
\vspace{-0.2in}
\end{table}

\begin{figure*}[t]
	\begin{minipage}[t]{0.61\linewidth}
    \centering
      \resizebox{0.95\textwidth}{!}{
      \begin{tabular}{c||ccc|ccc|ccc}
      \hline
      \multirow{2}{*}{Task} & \multicolumn{3}{c|}{Blacklight's $salt_Q$ off}
        & \multicolumn{3}{c|}{$salt_Q$ on}
        & \multicolumn{3}{c}{attacker knows ($\mathbf{p},\mathbf{q},\mathbf{w}$), $salt_Q$ on}                     \\ \cline{2-10}
                            & \multicolumn{1}{c}{Boundary} & \multicolumn{1}{c}{ECO}  & HSJA & \multicolumn{1}{c}{Boundary} & \multicolumn{1}{c}{ECO} & HSJA & \multicolumn{1}{c}{Boundary} & \multicolumn{1}{c}{ECO} & HSJA \\ \hline
      \mnist                 & \multicolumn{1}{c}{0\%}      & \multicolumn{1}{c}{0\%}  & 0\%  & \multicolumn{1}{c}{0\%}      & \multicolumn{1}{c}{0\%} & 0\%  & \multicolumn{1}{c}{0\%}      & \multicolumn{1}{c}{0\%} & 0\%  \\
      \gtsrb                 & \multicolumn{1}{c}{10\%}     & \multicolumn{1}{c}{5\%}  & 5\%  & \multicolumn{1}{c}{0\%}      & \multicolumn{1}{c}{0\%} & 0\%  & \multicolumn{1}{c}{0\%}      & \multicolumn{1}{c}{0\%} & 0\%  \\
      \cifar                 & \multicolumn{1}{c}{20\%}     & \multicolumn{1}{c}{15\%} & 25\% & \multicolumn{1}{c}{0\%}      & \multicolumn{1}{c}{0\%} & 0\%  & \multicolumn{1}{c}{0\%}      & \multicolumn{1}{c}{0\%} & 0\%  \\
      \imagenet              & \multicolumn{1}{c}{5\%}      & \multicolumn{1}{c}{10\%} & 20\% & \multicolumn{1}{c}{0\%}      & \multicolumn{1}{c}{0\%} & 0\%  & \multicolumn{1}{c}{0\%}      & \multicolumn{1}{c}{0\%} & 0\%  \\ \hline
      \end{tabular}}
\vspace{-0.1in}
  \captionof{table}{
\em Attack success rate using guided
  transformations attacks.}  \label{table:guided_trans}
  \end{minipage}
  \hfill
    \begin{minipage}[t]{0.38\linewidth}
    \centering
    \resizebox{0.8\textwidth}{!}{
    \begin{tabular}{c|cc}
    \hline
    Metrics   & NES & AutoZOOM \\ \hline
    Attack success \%    & 100\%    & 100\%    \\
    Attack detect \%    & 100\%    & 100\%    \\
    Detection coverage & 99.1\%    & 98.9\%  \\
    Avg queries to detection   & 2   & 2 \\
    Avg \# of attack queries & 1473    & 1240 \\ \hline
    \end{tabular}}\vspace{-0.1in}
    \captionof{table}{\em Blacklight vs. hybrid batch attacks.}
    \label{table:hybrid}
  \end{minipage}
  \vspace{-0.2in}
\end{figure*}

\para{Increasing Perturbation Budgets.} Our evaluation so far assumes
the attacker's
perturbation budget is limited to commonly accepted values: 0.05 for
both L$_\infty$ and {\em
  normalized} L$_2$. Future attacks might tolerate a higher perturbation budget in specific settings.  Thus, we
evaluate Blacklight's detection performance against attacks on \cifar{} with
larger perturbation budgets. For all attacks, we incrementally increase the budget from 0.05 all the
way up to 0.2, and measure Blacklight's attack detection rates for
each budget
level (running 20 attack instances for each data point). We show that
the attack detection
rates for NES variants and ECO remain steady at 100\%, but Boundary and HSJA begin to evade detection when {\em normalized}
L$_2 = 0.15$ (or L$_2 = 8.3$).

Blacklight can compensate by tuning the
fingerprint matching threshold $\mathbf{T}$. Table~\ref{table:perturb}
shows that by lowering $\mathbf{T}$ from 25 to 15, we can dramatically increase detection
rates, restoring perfect detection to most attacks (except HSJA at {\em
  normalized} L$_2 = 0.15/0.2$ (L$_2 = 8.3/11.1$) and Boundary at {\em
  normalized} L$_2 = 0.2$ (L$_2 = 11.1$)). This drop in $\mathbf{T}$ only increases
false positive rates by 0.74\%.

We further validate our results on the other three tasks for the two fastest
converging attacks (ECO and HSJA) and the results
(Table~\ref{table:perturb_all_models}) are consistent with
\cifar. Finally, we
also perform analysis on the L$_2$ distances between benign images to provide
a baseline for reasonable L$_2$ budget for adversarial attacks in Appendix
\S\ref{sec:additional_results_increase_purb}.

\para{Evasion via Guided Transformation.}
Beyond first order adaptive attacks, we worked hard to design more
powerful attacks specifically targeting Blacklight.  \htedit{Assuming a Blacklight system's
parameters $\mathbf{q}$ and $\mathbf{w}$ are unknown to an attacker},  the strongest
attack we could design is the two-pronged reverse engineer attack, where an
attacker first uses queries to probe the limits of $\mathbf{q}$ and
$\mathbf{w}$, and then leverages those results to optimize a guided
transformation attack.


The high-level intuition is that an attacker can optimally spread out their
perturbation budget across the image, if they understand Blacklight and
learned its specific configuration parameters. As long as there is at least one
pixel changed (after pixel quantization) for some sliding window, hash values
of the window will be changed. Thus, the attacker just needs to make sure
that for each window, at least one pixel is different from all prior
queries after quantization. In this case, Blacklight's use of $salt_Q$
in eq (\ref{eq:quantization}) is crucial to
resisting these guided transformation attacks. Next, we summarize the attack and
results when Blacklight turns $salt_Q$ off or on.

\spara{Guided transformation (Blacklight's $salt_Q$ off).} An attacker begins by
estimating quantization step $\mathbf{q}$ and using it to compute quantization
boundary $B$, followed by estimating value of $\mathbf{w}$. It does this by
issuing pairs of queries with a minimal perturbation based on an
initial estimate of $\mathbf{q}$ or $\mathbf{w}$, and observing whether the second query is
detected as an attack. This is repeated using binary search until both
$\mathbf{q}$ and $\mathbf{w}$ are determined.  Finally, the attacker computes $B$ from
$\mathbf{q}$, and then the optimal layout of modified pixels
to maximize the number of substring windows affected by the perturbation.
The attacker uses this process to modify each query to evade
detection while iteratively optimizing queries to generate the adversarial
example.  We implement this attack on top of the two fastest converging
attacks (ECO and HSJA) and the slowest attack (Boundary).
Table~\ref{table:guided_trans} shows that the
attacker achieves no more than $25\%$ success rate for all
tasks.

\spara{Guided transformations (Blacklight's $salt_Q$ on).}  The defender can overcome the
above adversary by making it harder to extract the quantization
boundary. Blacklight does so by adding a ``salt'' to the quantization
process, i.e., $salt_Q$ in eq. (\ref{eq:quantization}).
This
defeats attempts by the attacker to reverse engineer $\mathbf{q}$ and $B$.  Without knowledge of $\mathbf{q}$, an attacker can still launch a weaker version of
the attack, but must overshoot on perturbation to increase chances of it
persisting through the salted quantization and alter the hashes.  We implement such
attack by altering $5$, $10$, and $15$ out of every $20$ pixels within the
perturbation budget. When applying this new attack on top of ECO,
HSJA, and Boundary, the attacker still achieves 0\% success on all tasks,
\htedit{while Blacklight maintains a high detection coverage (78\%).}  This confirms the significant robustness
gained by adding the salt.

\htedit{
  \para{Guided Transformations when Attacker Knows $(\mathbf{q},
    \mathbf{p},\mathbf{w})$.}  Finally, we
  consider the strongest guided transformation attack -- the
  attacker knows the exact values of $\mathbf{q}$, $\mathbf{p}$,
  $\mathbf{w}$ and can better perturb queries to evade detection.

 To make a query $x$ evade detection, the
  attacker must ensure that for each window, at least one pixel of $x$ is different from all
  prior queries after quantization. This is because
  Blacklight's one-way hash distribution and the top
  $\mathbf{S}$ hash choices remain unpredictable to the attacker.
  Knowing $\mathbf{q}$, $\mathbf{p}$,
  $\mathbf{w}$ helps the attacker to  optimize the pixel perturbation.
  For example, now in each window changing a
  pixel by $\mathbf{q}$ or $-\mathbf{q}$ will change the hash despite the use of
  $saltQ$.  To make $x$'s full hashes different from
  those of all prior attack queries, we apply a permutation-based
  pixel selection
  algorithm to
  minimize the total perturbation (see Algorithm\ref{alg:guided_trans_alg} in
  Appendix).

  Even with this strong attack, attackers still have 0\% success rate after
  sending 100K queries (see
  Table~\ref{table:guided_trans}). These attack queries do bypass
  Blacklight's detection, but the attack's iteration optimization
  process never converges to generate an adversarial example
  (regardless of the perturbation budget).  This is because
  the perturbation applied to individual attack queries in order to evade detection is
  too large to make the query results useful for attack optimization,  i.e.,
  they fail to capture detailed decision boundaries of the
  target model.  As such, the iterative optimization process fails to
  make concrete progress but ``randomly'' wanders around.

  Together, our experiments with guided transformation attacks show
  that (1) salted quantization is important to resist advanced attackers,
  and (2) under the Blacklight defense, attackers now face two
  conflicting goals when building attack queries: evading Blacklight's
  detection or advancing the attack's iterative optimization process
  using queries.
}

\vspace{-0.1in}
\subsection{Reducing Number of Attack Queries}
\label{subsec:reduce_queries}
\vspace{-0.1in}
Another way to evade Blacklight is to reduce the queries needed
for an attack to succeed. Since Blacklight examines similarity between
a new query and past queries, the fewer the queries  needed,
the lower the probability that the attack query will be detected.
We explore two adaptive attacks that focus on reducing attack queries needed.


\para{Hybrid Black-Box Attacks.}
\htedit{Substitute model based priors can be useful for planning
  attack queries~\cite{suya2020hybrid, juuti2019making,
    huang2019black, cheng2019improving}. For example, adversarial
examples generated from a substitute model can serve as a good
starting point to launch query-based black-box attacks, allowing the attacker to
use less number of queries to complete the
attack~\cite{suya2020hybrid}. }
We run two of these hybrid attacks~\cite{suya2020hybrid} (NES and AutoZOOM) while using Blacklight to protect
the target model. For each attack, we run 100 attack sequences
on \cifar{} and report our results in
Table~\ref{table:hybrid}.
We see that the two hybrid attacks do reduce the number of queries
required for complete an attack, Blacklight still leads to 100\%
attack detection,  99\% of detection coverage, and detect attack
queries after just 2 queries.


\para{Optimal Black-Box Attacks.}
Since black-box attacks are
continuously evolving in query efficiency, we also evaluate Blacklight
against two types of highly efficient attacks that are possible but do not
yet exist. First, we consider extremely ``query-efficient'' black-box attacks
that require orders of magnitude fewer attack queries than current
attacks by downsampling existing attack sequences. We find that even when attacks are able to
complete in 500, 100, or 50 queries, Blacklight still detects them near
perfectly (100\% detection rate for 4 attacks and 89\% for Boundary attack).

Second, we imagine a ``perfect-gradient'' black-box algorithm that is somehow
able to perfectly predict gradient functions from the results of its attack
queries, as accurately as a white-box attack. Our results show
Blacklight detects $100\%$ of attacks driven by
CW~\cite{carlini2017towards}, and $81\%$ of attacks driven by
PGD~\cite{madry2017towards}. The details are listed in
Table~\ref{table:stronger_attack}, Appendix \S\ref{sec:opt}.

\vspace{-0.1in}
\subsection{Evasion by Exploiting Reset Window}
\vspace{-0.05in}
\label{sec:pause}
Finally, to guarantee the efficacy of Blacklight, the defender would
reset the system periodically. Thus, a patient attacker can leverage
the reset feature to evade detection.

\para{Pause and Resume Attacks.}  Adversaries can try to evade detection by
exploiting the fact that Blacklight periodically resets its database
to remove all fingerprints.  They
can pause their attack every time it receives a
rejection response, and resuming the attack the next time Blacklight resets
its database. We experiment on all five black-box attacks using this
strategy against a \cifar{} model and Blacklight. We run 100 instances of
each attack, and show average total queries needed for each attack to
succeed, and the average number of reset cycles that requires in
Table~\ref{table:cycles}.  If we reset Blacklight every 24 hours, the fastest
successful attacker would complete an attack (using HSJA) in $1092$ days or
roughly 3 years. While this strategy does allow for a successful attack, the
time cost to perform this attack makes it highly impractical.

\vspace{-0.15in}
\htedit{
\section{Conclusion and Limitations}
\vspace{-0.1in}
\label{sec:discussion}

Blacklight protects DNN models against query-based black-box attacks,
using a probabilistic fingerprint to detect highly similar queries
generated by attack optimization. Blacklight achieves
near-perfect detection against eight SOTA attacks with negligible
false positives, resists persistent attackers, and is
robust to a range of adaptive and even idealized
countermeasures. We also demonstrated that Blacklight can
successfully generalize to some text classification tasks.

Blacklight faces two limitations that demand further research. First, it is
unable to defend against substitute model (SM) attacks, but can be combined
with SM defenses to launch a more complete defense against both types of
black-box attacks (see Appendix \S\ref{sec:hybrid_submodel} for initial
results). Second, Blacklight relies on the fact that {\em existing}
query-based black-box attacks all produce highly similar queries during their
iterative optimization process, a phenomenon rarely seen in benign
queries. It is not future-proof, i.e. a (future) attack breaking this assumption
would evade Blacklight. }

{
 \footnotesize
 \bibliographystyle{acm}
 \bibliography{zhao,ms}
}

\clearpage
\appendix

\section*{Appendix}
\label{sec:appendix}

\htedit{This appendix consists of the following
items:
\begin{packed_itemize}

\item   \S\ref{sec:theory_app} presents the full detail of our
  formal analysis mentioned in \S\ref{sec:theory}, including the key
  theoretical result and its proof and how these results can be used
  to guide Blacklight's parameter configuration;

\item \S\ref{sec:hybrid_submodel} describes how ensemble adversarial
  training can be combined with Blacklight as a hybrid defense against
  both substitute model attacks and query-based black-box attacks
  (mentioned in \S\ref{sec:backoverview});
\item \S\ref{sec:existing_app} is a supplement of \S\ref{sec:existing}
  by providing detailed evaluation of SD and
  PRADA under a persistent attacker, who switches to a new account
  when the current account is detected as adversarial and thus
  banned.

\item \S\ref{sec:design_app} show the results where we empirically verify the two assumptions
we made in \S\ref{sec:design}.

\item \S\ref{sec:extra_config} summarizes the experimental
  configurations used by our experiments, including classification tasks, datasets, model training configurations,
model architectures, and attack perturbation budgets, as well as
Blacklight's configuration.

\item \S\ref{sec:eval_app} includes additional results for
  Blacklight's performance on detection and mitigating Boundary Attacks with 1 million query limit (mentioned in \S\ref{subsec:8.2}),
 universal adversarial patches (discussed in \S\ref{sec:uni_patch}) and the impact for Blacklight
parameter settings (discussed in \S\ref{sec:param_analysis}).

\item \S\ref{sec:additional_results} provides detailed discussion and experimental results
on adaptive attacks discussed in \S\ref{sec:countermeasure},
including 5 subsections: Evasion via Image Transformations,
Increasing Perturbation Budget,
Guided Transformations when Attacker Knows ($\mathbf{q},\mathbf{p},\mathbf{w}$),
Optimal Black-Box Attacks and Pause and Resume Attacks.

\end{packed_itemize}
}

 \vspace{-0.1in}
\section{Formal Analysis of Blacklight}
\vspace{-0.1in}
\label{sec:theory_app}
\label{subsec:theorem}
We formally examine Blacklight by
modeling its process of probabilistic fingerprinting. We derive analytical bounds on the probability of Blacklight
flagging
a query pair $(x,y)$ as attacks, and subsequently estimate
Blacklight's false positive rate and attack query detection coverage.

\vspace{-0.1in}
\subsection{Definitions}
\label{subsec:thdef}
\htedit{We first introduce the terms that we will use to model the proposed
probabilistic fingerprinting process on input queries.}

\begin{definition}  \label{def:hash} {\bf Hash Function} is a function that, for a given input $x$, produces $\mathbf{N}$ hash values,  $\mathbf{H}_x=(h_{1},h_{2}, ...,
  h_{\mathbf{N}})$.  Each entry $h_i$ is a positive integer that is independent and identically
  distributed (I.I.D.) in the hash space $[1,\Omega]$, where $\Omega$
  is a very large positive integer,
  $\Omega >>\mathbf{N}$.   Without loss of
  generality, $h_i$ follows a uniform distribution within
  $[1,\Omega]$.
  \end{definition}

\begin{definition} Given two queries $x$ and $y$,  we represent their full
hash set as $ \mathbf{H}_x= \mathbf{H}_{sh} \cup \hat{\mathbf{H}}_{x}$
and $\mathbf{H}_y= \mathbf{H}_{sh} \cup \hat{\mathbf{H}}_{y}$,
where $\mathbf{H}_{sh}=\mathbf{H}_x \cap \mathbf{H}_y$,
$\hat{\mathbf{H}}_x \cap \hat{\mathbf{H}}_y=\emptyset$,
  $|\mathbf{H}_{sh}|= N-\Delta$,
  $|\hat{\mathbf{H}}_{x}|=|\hat{\mathbf{H}}_{y}|=\Delta$. For simplicity,
  we  assume $\mathbf{H}_{sh}  \cap
  \hat{\mathbf{H}}_x=\emptyset$, $\mathbf{H}_{sh}  \cap
  \hat{\mathbf{H}}_y=\emptyset$.
  \label{def:hashdiff}
\end{definition}
\noindent Note that $\Delta$ represents the amount of full hash differences
between $x$ and $y$. \htedit{We also empirically
  validate the assumption of $|\mathbf{H}_{sh}  \cap
  \hat{\mathbf{H}}_x|/|\mathbf{H}_x|\approx 0$  on \cifar{}}.

\begin{definition}
  {\bf Probabilistic Fingerprinting (PF)} is a function performed on the
  full hash set
 that samples top $S$ hash entries out of
  $\mathbf{H}_x$, {\em i.e.\/}  $\mathbb{S}(\mathbf{H}_x)=(h'_1,h'_2,...,h'_S) \subset
  \mathbf{H}_x$.
\end{definition}
\noindent Finally, Blacklight operates on $\mathbb{S}(\mathbf{H}_x)$
  to detect attack queries rather than the full hash set
  $\mathbf{H}_x$.  Blacklight marks $(x,y)$ as attack
images if $| \mathbb{S}(\mathbf{H}_x) \cap
\mathbb{S}(\mathbf{H}_y)| > \mathbf{T}$.

\vspace{-0.1in}
\subsection{Key Results}
\label{subsec:thresult}

Our analysis led to the following theorem.
\begin{theorem}\vspace{-0.05in}
Let  $Q(\Delta)$ be the probability
of Blacklight flagging a query pair $(x,y)$ as attack
queries where $x$ and $y$'s full hashes differ by $\Delta$ entries.
Then $Q(\Delta) \leq Q^{upper}(\Delta)$.
{\small
\begin{align}  \vspace{-0.1in}
  \begin{split}
 & Q(\Delta) \triangleq   Pr \left(
                          \text{Blacklight}\;(x,y)=\text{attack}\; |
                          \;\text{diff}\; (\mathbf{H}_x,
                          \mathbf{H}_y)=\Delta \right)  \\
  & Q^{upper}(\Delta)  =
       \sum_{k=\mathbf{T}+1}^{\min(\mathbf{S}, \mathbf{N}-\Delta)} {\binom{\mathbf{N}-\Delta}{k} \cdot
         \binom{\Delta}{\mathbf{S}-k}}/{\binom{\mathbf{N}}{\mathbf{S}}}
    \label{eq:upperbound}
 \end{split}
\vspace{-0.05in}
\end{align}
}

where $\mathbf{N}$, $\mathbf{S}$ and $\mathbf{T}$ are
parameters of Blacklight (see \S\ref{sec:design}.).
\label{th:detection}
\end{theorem}  \vspace{-0.05in}

\begin{proof}
Clearly $\mathbb{S}(\mathbf{H}_x)=\mathbb{S}(\mathbf{H}_{sh} \cup \hat{\mathbf{H}}_{x})$ will contain
entries from $\mathbf{H}_{sh}$ and $\hat{\mathbf{H}}_{x}$. The
same applies to $\mathbb{S}(\mathbf{H}_y)$.  Since $\hat{\mathbf{H}}_x \cap
  \hat{\mathbf{H}}_y=\emptyset$, the overlapping entries of
$\mathbb{S}(\mathbf{H}_x)$ and $\mathbb{S}(\mathbf{H}_y)$ will only come
from $\mathbf{H}_{sh}$. That is,
\begin{eqnarray}
 && (\mathbb{S}(\mathbf{H}_x) \cap \mathbb{S}(\mathbf{H}_y)) \subset
    \mathbf{H}_{sh}, \\
&&  (\mathbb{S}(\mathbf{H}_x) \setminus
   \mathbb{S}(\mathbf{H}_y))\subset \hat{\mathbf{H}}_{x} \\
  &&  (\mathbb{S}(\mathbf{H}_y) \setminus \mathbb{S}(\mathbf{H}_x)) \subset \hat{\mathbf{H}}_{y}
  \end{eqnarray}
To calculate the upper bound on $Pr(|\mathbb{S}(\mathbf{H}_x) \cap \mathbb{S}(\mathbf{H}_y)| >
\mathbf{T})$, we consider the ``optimal scenario'' using a custom-designed\footnote{ One
possible design is picking hash entries by their indices.  If the fingerprinting process
chooses the same set of hash indices for $x$ and $y$,   the chosen entries in
$\mathbf{H}_{sh}$ will be the same for $x$ and $y$.} probabilistic
fingerprinting process, so that when picking entries from
$\mathbf{H}_x$ and $\mathbf{H}_y$,  the chosen entries in
$\mathbf{H}_{sh}$ are always the same for $x$ and $y$.  This is to maximize the similarity between
$\mathbb{S}(\mathbf{H}_x)$ and $\mathbb{S}(\mathbf{H}_y)$, which will
be higher than that offered by selecting top $\mathbf{S}$ entries.
Thus we compute the upper bound  as the probability of more than $\mathbf{T}$ entries in $\mathbb{S}(\mathbf{H}_x)$
  (and $\mathbb{S}(\mathbf{H}_y)$) come from
  $\mathbf{H}_{sh}$ and the rest come from $\hat{\mathbf{H}}_{x}$
  ($\hat{\mathbf{H}}_{y}$).
 Since each hash entry's value is i.i.d.,  and $|\mathbf{H}_{sh}|=\mathbf{N}-\Delta$, $|\hat{\mathbf{H}}_{x}|=\Delta$, we
  calculate the probability following the hypergeometric
  distribution and arrive at  the upper bound shown in the theorem.
\end{proof}

\begin{figure}[h]
  \centering
  \includegraphics[width=0.38\textwidth]{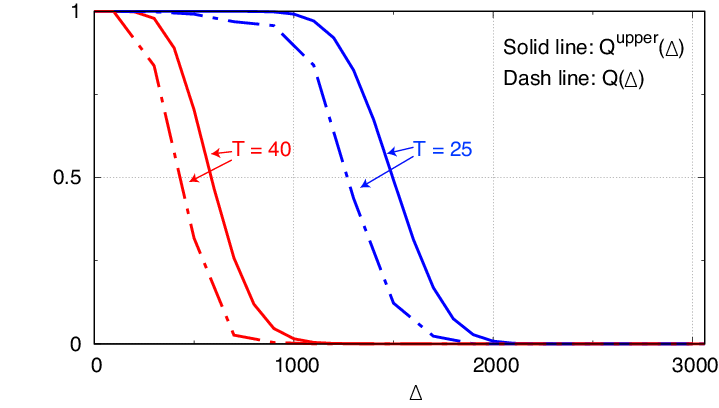}
  \vspace{-0.15in}
  \caption{\em Measured $Q(\Delta)$ and its theoretical
    upper-bound $Q^{upper}(\Delta)$, both decaying
    fast with $\Delta$. The results are for \cifar~ queries
    ($\mathbf{N}=3053, \mathbf{S}=50, \mathbf{T}=25$ or $40$).}
    \vspace{-0.0in}
  \label{fig:detection_theory}
\end{figure}

\para{Key Observation: $Q^{upper}(\Delta)$ Decaying Fast with $\Delta$.}
While unable to simplify its symbolic expression, we empirically found that $Q^{upper}(\Delta)$ can be approximated by a
symmetrical sigmoidal function of $\Delta$ (with the goodness of fit
$R^2$=0.9996). For instance, consider two configurations that Blacklight uses to scan \cifar\
image queries:  $\mathbf{N}$ = 3053 ($w$ = 20, $p$ = 1), $\mathbf{S}=50$,  and  $\mathbf{T}=$25 or
40. Then $ Q^{upper}(\Delta)$ can be approximated as:
{\small
\begin{equation*}
Q^{upper}(\Delta)\approx\left\{\begin{matrix}
1.011 \cdot \left(1 +
    (\frac{\Delta}{1494.85})^{11.77}\right)^{-1}-0.013,  &\mathbf{T}=25 \\
1.006 \cdot \left(1 +
    (\frac{\Delta}{584.51})^{5.97}\right)^{-1}-0.009,  & \mathbf{T}=40\\
\end{matrix}\right.
\end{equation*}
}
\noindent Note that we followed the standard curve fitting process to approximate $Q^{upper}(\Delta)$,
and $1.011/1.006$ are function parameters generated by curve fitting.
For the above configurations, Figure~\ref{fig:detection_theory}
plots the upper bound  $Q^{upper}(\Delta)$ as a function of $\Delta$ and also $Q(\Delta)$ measured  by
running Blacklight on both benign and attack queries generated from
\cifar.
We
see that the upper bound is reasonably tight.  More importantly,
both decay very fast with
$\Delta$.

\para{Blacklight's Detection Coverage \& False Positive Rate.}
We model Blacklight's two performance metrics from $Q(\Delta)$:
\begin{align*} \vspace{-0.05in}
 & \text{False positive rate} \leq Q(\Delta_{benign}) \\
  & \text{Attack query detection coverage} \geq Q(\Delta_{attack}) \vspace{-0.05in}
\end{align*}
where the attack query detection coverage is the probability of detecting a pair
of attack queries as adversarial, and the false positive rate is the
probability of detecting a pair of benign queries as adversarial.  And
$\Delta_{benign}$ is the minimum full hash difference between
benign queries, and $\Delta_{attack}$ is the maximum full hash difference between
attack queries. Since $Q(\Delta)$ decays fast with
$\Delta$, a properly configured Blacklight system can effectively
detect attack queries at a low false
positive rate, i.e., $Q(\Delta_{attack})\rightarrow 1,
\;Q(\Delta_{benign})\rightarrow 0$,  as long as $\Delta_{benign} >>\Delta_{attack}$.


\htedit{
\vspace{-0.1in}
\subsection{Guiding the Parameter Configuration}
\label{subsec:thguide}


Our analysis shows that Blacklight's system parameters will impact
${Q^{upper}}(.)$ and thus its false positive rate and
detection coverage. We leverage this modeled relationship to
guide Blacklight's parameter configuration. The goal is to meet a
desired false positive while maximizing detection coverage.

\para{Choosing $\mathbf{T}$.} Among the five parameters,  $\mathbf{T}$
is of particular importance since
it defines the threshold of fingerprint
matching. Figure~\ref{fig:detection_theory} already shows that
$\mathbf{T}$ can largely alter the range of ${Q^{upper}}(.)$. To
select $\mathbf{T}$, we propose to examine the model's training
data to compute $\Delta_{benign}$ and use it to choose $\mathbf{T}$
to meet a desired false positive rate. For example,  when
$\mathbf{N}=(|x|-\mathbf{w+p)/p}$=3053,  the full hashes of quantized images in \cifar\ produce
$\Delta_{benign}$=2638.  Thus if  $\mathbf{S}$=50, then
$\mathbf{T}=mathbf{S}/2=25$ should produce a reasonably small false positive rate
while maintaining a high detection rate.

%

\para{Choosing  $\mathbf{w}$, $\mathbf{p}$, $\mathbf{q}$,
  $\mathbf{S}$.} We divide these parameters into three groups: [$\mathbf{w}$, $\mathbf{p}$],  [$\mathbf{q}$],  and [$\mathbf{S}$].
Group one decides how to map a query into a full
set of $\mathbf{N}$ hashes, and how many hashes a single pixel could
affect.  The second group ($\mathbf{q}$), together with the first group,
controls the full hash similarity among
attack and benign queries, i.e., $\Delta_{attack}$ and $\Delta_{benign}$ defined by
our formal analysis.  Finally, $\mathbf{S}$ (and $\mathbf{T}$)
determine how to compare queries' similarity by their hashes.  With
these in mind, we propose the following guidelines.


We should choose $\mathbf{q}$ as a moderate value to make attack
queries' hashes highly similar (i.e., small $\Delta_{attack}$ thus
high detection coverage $Q(\Delta_{attack})$), but
not too large to diminish the difference between benign queries (i.e.,
large $\Delta_{benign}$ to maintain a low false
positive rate approximated by
$Q(\Delta_{benign})$).

 $\mathbf{S}$ should be much less than $\mathbf{N}$ for  scalability,
yet not too small so the fingerprint has enough capacity to capture
the difference between benign queries,  thus keeping $\Delta_{benign}$
sufficiently large to maintain a low false positive rate
$Q(\Delta_{benign})$.

The choice of $\mathbf{w}$ could affect both false positive rate and detection coverage.  The larger the
$\mathbf{w}$, the more hashes that changing one pixel will
affect, and more sensitive the  fingerprint will react to content
variation, thus increasing $\Delta_{attack}$
and $\Delta_{benign}$.   As such, increasing $w$ will reduce both
false positive $Q(\Delta_{benign})$ and detection coverage
$Q(\Delta_{attack})$.   Ideally, one should choose the smallest $w$
that meets the desired false positive rate.

}

\section{Hybrid Defense against the Substitute Model Attack}
\label{sec:hybrid_submodel}
Blacklight is designed to detect query based
black-box attacks. It  cannot defend against attacks transferred from
a substitute model. As we discussed in \S\ref{sec:back}, substitute
model attacks can be effectively stalled by an existing defense called
ensemble adversarial training (EAT)~\cite{tramer2018ensemble}. EAT
adversarially trains an ensemble of models with different
architectures~\cite{madry2017towards}, which are shown to be robust against the substitute model
attack. Hence, to defend against all types of black-box
attacks, the defender can combine Blacklight with EAT to build a
hybrid defense system.

We build and evaluate a hybrid Blacklight and EAT defense on the
cifar{} task.  Specifically, we build an ensemble model with three
different architectures (6-layer CNN, 8-layer CNN, ResNet-20) and
adversarially train the network using PGD attacks as suggested
by~\cite{madry2017towards}.  We use the same Blacklight configuration
as before.

We perform both substitute model based attacks and query based black-box
attacks against the above ensemble model defended by Blacklight. For
the substitute model attack we run the state-of-art attack proposed by Papernot et
al~\cite{papernot2017practical}, and for the query-based attacks we
run the same five black-box attacks.  The result shows that the hybrid
defense works well and the two defenses do not interfere with each
other.  The substitute model attack achieves 0\% success (thanks to
EAT), and Blacklight achieves the same accurate attack query detection as
reported before. Thus, we conclude that Blacklight, when combined with
EAT,  can defend against today's black-box attacks.

\vspace{-0.08in}
\section{Additional Results for \S\ref{sec:existing}}
\label{sec:existing_app}

We show the detection performance of SD and PRADA under the assumption where attackers
will switch to a new account when the current account is detected as malicious
and banned in Table~\ref{table:compare}.

\begin{table}[t]
  \centering
  \resizebox{0.49\textwidth}{!}{
\begin{tabular}{l|cccc}
 \hline
 Attack & \begin{tabular}[c]{@{}c@{}}Detection\\coverage\end{tabular} &
 \begin{tabular}[c]{@{}c@{}}Avg queries \\to
   detect \end{tabular} & \begin{tabular}[c]{@{}c@{}}Attack
                              success \\ w. mitigation \end{tabular}
                        & \begin{tabular}[c]{@{}c@{}}Attack
                              success \\ w/o mitigation \end{tabular} \\
 \hline
 NES - QL & 1.8\%  / 0.8\% & 52 / 112 & 97\%   / 97\% & 97\%   \\
 NES - LO & 1.3\%   / 0.9\% & 52 / 111 & 85\% / 85\% & 85\% \\
 Boundary & 1.0\%  / 0.8\% & 54   / 115 & 86\%  / 86\%  & 86\% \\
 ECO     & 1.8\%   / 0.9\% & 53 / 112 & 88\%   / 88\%  & 88\%  \\
 HSJA   & 1.7\%  / 0.9\% & 52 / 111 & 100\%  / 100\% & 100\%  \\
 QEBA   & 1.6\%  / 0.9\% & 52 / 111 & 100\%  / 100\% & 100\%  \\
 SurFree   & 1.9\%  / 0.9\% & 52 / 111 & 100\%  / 100\% & 100\% \\
 Policy-Driven   & 2.1\%  / 0.9\% & 53 / 111 & 98\%  / 98\% & 98\%  \\
 \hline
\end{tabular}}
\captionof{table}{\em Detection performance of Stateful Detection~\cite{chen2020stateful} and
 PRADA~\cite{juuti2019prada} when attackers change their accounts after detected and disabled on \cifar{},
 in terms of attack detection and mitigation. The result is presented as  s``Stateful Detection
 / PRADA''.} \vspace{-0.1in}
 \label{table:compare}
\end{table}

\vspace{-0.08in}
\section{Additional Results for \S\ref{sec:design}}
\label{sec:design_app}

\begin{figure}[t]
   \centering
   \includegraphics[width=.32\textwidth]{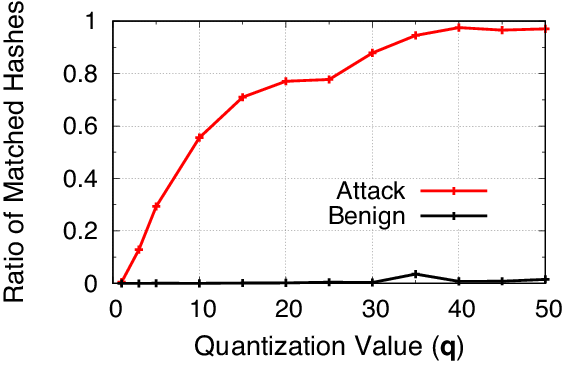}
     \vspace{-0.05in}
     \caption{\em Average ratio of matched hashes in
         fingerprints of 10,000 pairs of quantized attack queries and
         10,000 pairs of  quantized benign queries, all for \cifar{}, when varying the
         quantization step ($\mathbf{q}$). We can see that
     quantization does increase the similarity between attack query fingerprints but have 'negligible' impact
     on beign query fingerprints.}
   \label{fig:quantization_match_comp}
 \end{figure}

\begin{figure}[t]
    \centering
    \includegraphics[width=.35\textwidth]{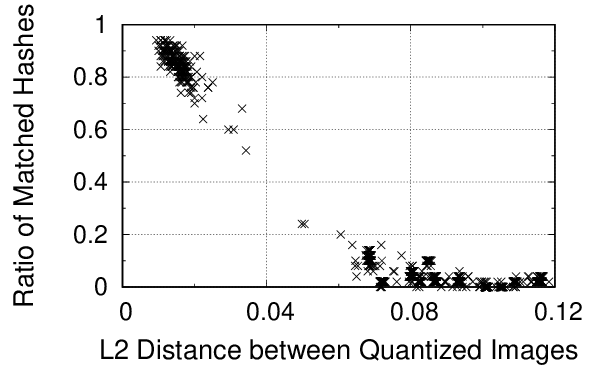}
      \vspace{-0.05in}
      \caption{\em \htedit{We empirically demonstrate that highly similar
          image queries (after quantization) also have highly similar
         fingerprints, based on 10000 pairs of attack queries on
         \cifar. In x-axis we plot the $L_2$ distance between a pair
         of attack queries after they are quantized, and in y-axis, we
         plot the ratio of matching in their probabilistic
         fingerprints. We see that the two metrics are strongly
         (negatively) correlated. }}
     \label{fig:l2_match}
     \vspace{-0.1in}
  \end{figure}

We empirically validate two assumptions we make in \S\ref{sec:design}.
\begin{packed_itemize}
  \item Quantization increases the similarity between attack queries. We empirically
  validate it by showing the average number of matched hashes in fingerprints of attack/benign
  queries with different quantization step ($\mathbf{q}$). As shown in Figure~\ref{fig:quantization_match_comp},
  quantization not only increase the similarity between attack queries, but also have
  little impact on benign queries, which is ideal for attack detection.
  \item Highly similar (quantized) queries will have highly similar fingerprints.
  We empirically verify this with Figure~\ref{fig:l2_match}. We can see that the images
  with smaller $L_2$ distances have a higher ratio of hashes match in their fingerprints.
\end{packed_itemize}

\vspace{-0.15in}
\section{Experimental Configurations}
\label{sec:extra_config}
\vspace{-0.08in}
\subsection{Classification Tasks and Models}
\vspace{-0.08in}
Table~\ref{table:tasks} summarizes the
four image classification tasks that we use for our experiments.  Their associated models are listed
below:
\begin{packed_itemize}
\vspace{-0.08in}
\item {\bf \mnist{}} (Table~\ref{table:model_mnist}) is a convolutional
  neural network (CNN) consisting of two pairs of
convolutional layers connected by max pooling layers, followed by two
fully connected layers.

\item {\bf \gtsrb{}} (Table~\ref{table:model_gtsrb}) is a
CNN consisting of three pairs of
convolutional layers connected by max pooling layers, followed by two
fully connected layers.

\item {\bf \cifar{}} is a ResNet-20~\cite{he2016identity} that
includes $20$ sequential convolutional layers, followed by pooling, dropout, and fully connected layers.

\item {\bf \imagenet{}} is the ResNet-152~\cite{he2016deep} model trained on the ImageNet dataset~\cite{ILSVRC15}. It has $152$ residual blocks with over $60$ million parameters.

\end{packed_itemize}

\begin{table*}[t]
  \centering
  \resizebox{0.95\textwidth}{!}{
\begin{tabular}{l|lllllll}
\hline
Task                                          & Dataset  & \# Classes & \begin{tabular}[c]{@{}l@{}}Training \\ data size\end{tabular} & \begin{tabular}[c]{@{}l@{}}Test data\\ size\end{tabular} & Input size    & Model architecture & Model accuracy\\ \hline
Digit Recognition (\mnist) & MNIST    & 10         & 60,000                                                        & 10,000                                                   & (28, 28, 1)   & 6 Conv + 3 Dense   & 99.36\%\\
Traffic Sign Recognition (\gtsrb) & GTSRB    & 43         & 39,209                                                        & 12,630                                                   & (48, 48, 3)   & 6 Conv + 3 Dense   & 97.59\% \\
Object Recognition - Small (\cifar) & CIFAR-10 & 10         & 50,000                                                        & 10,000                                                   & (32, 32, 3)   & ResNet20           & 91.48\% \\
Object Recognition - Large (\imagenet) & ImageNet & 1000       & 1,281,167                                                     & 50,000                                                   & (224, 224, 3) & ResNet152        & 73.05\%  \\ \hline
\end{tabular}}
 \caption{\em Overview of image classification tasks with their associated datasets and models.}
 \label{table:tasks}
\end{table*}

\begin{table}[t]
 \centering
 \resizebox{0.49\textwidth}{!}{
\begin{tabular}{cccccccc}
\hline
Layer Index  &  Layer Name  &  Layer Type  &  \# of Channels  &  Filter Size  &  Activation   &  Connected to \\ \hline
1           &  conv\_1      &  Conv        & 32             & 3$\times$ 3          &  ReLU        &               \\
2           &  conv\_2      &  Conv        & 32             & 3$\times$ 3          &  ReLU        &  conv\_1       \\
2           &  pool\_1      &  MaxPool     & 32             & 2$\times$ 2         &  -           &  conv\_2       \\
3           &  conv\_3      &  Conv        & 64             & 3$\times$ 3          &  ReLU        &  pool\_1       \\
4           &  conv\_4      &  Conv        & 64             & 3$\times$ 3          &  ReLU        &  conv\_3       \\
4           &  pool\_2      &  MaxPool     & 64             & 2$\times$ 2         &  -           &  conv\_4       \\
5           &  conv\_5      &  Conv        & 128            & 3$\times$ 3          &  ReLU        &  pool\_2       \\
6           &  conv\_6      &  Conv        & 128            & 3$\times$ 3          &  ReLU        &  conv\_5       \\
6           &  pool\_3      &  MaxPool     & 128            & 2$\times$ 2         &  -           &  conv\_6       \\
7           &  fc\_1        &  FC          & 512            &  -            &  ReLU        &  pool\_3       \\
8           &  fc\_2        &  FC          & 512             &  -            &  ReLU       &  fc\_1         \\
8           &  fc\_3        &  FC          & 10              &  -            &  Softmax     &  fc\_2         \\ \hline
\end{tabular}
}
\caption{\em Model Architecture for \mnist.}
\label{table:model_mnist}
\end{table}

\begin{table}[t]
 \centering
 \resizebox{0.49\textwidth}{!}{
\begin{tabular}{cccccccc}
\hline
Layer Index  &  Layer Name  &  Layer Type  &  \# of Channels  &  Filter Size  &  Activation   &  Connected to \\ \hline
1           &  conv\_1      &  Conv        & 32             & 3$\times$ 3          &  ReLU        &               \\
2           &  conv\_2      &  Conv        & 32             & 3$\times$ 3          &  ReLU        &  conv\_1       \\
2           &  pool\_1      &  MaxPool     & 32             & 2$\times$ 2         &  -           &  conv\_2       \\
3           &  conv\_3      &  Conv        & 64             & 3$\times$ 3          &  ReLU        &  pool\_1       \\
4           &  conv\_4      &  Conv        & 64             & 3$\times$ 3          &  ReLU        &  conv\_3       \\
4           &  pool\_2      &  MaxPool     & 64             & 2$\times$ 2         &  -           &  conv\_4       \\
5           &  conv\_5      &  Conv        & 128            & 3$\times$ 3          &  ReLU        &  pool\_2       \\
6           &  conv\_6      &  Conv        & 128            & 3$\times$ 3          &  ReLU        &  conv\_5       \\
6           &  pool\_3      &  MaxPool     & 128            & 2$\times$ 2         &  -           &  conv\_6       \\
7           &  fc\_1        &  FC          & 512            &  -            &  ReLU        &  pool\_3       \\
8           &  fc\_2        &  FC          & 512             &  -            &  ReLU       &  fc\_1         \\
8           &  fc\_3        &  FC          & 43             &  -            &  Softmax     &  fc\_2         \\ \hline
\end{tabular}
}
\caption{\em Model Architecture for \gtsrb.}
\label{table:model_gtsrb}
\end{table}

\begin{table}
 \resizebox{.48\textwidth}{!}{
 \begin{tabular}{l|l}
 \hline
 Model & Training Configuration \\ \hline
 \mnist & epochs=50, batch=128, optimizer=Adam, lr=0.001 \\ \hline
 \gtsrb & epochs=50, batch=128, optimizer=Adam, lr=0.001 \\ \hline
 \cifar & \begin{tabular}[c]{@{}l@{}}epochs=200, batch=32, optimizer=Adam, lr=0.001\\(learning rate reduced after 80, 120, 160, 180 epochs)\end{tabular}\\ \hline
 \imagenet& Model trained and shared by He \etal~\cite{he2016deep} \\ \hline
 \end{tabular}}
 \caption{\em Detailed information on model training configurations for image classification tasks.}
 \label{tab:train_detail}
\end{table}

\vspace{-0.1in}
\subsection{Black-box Attack and Blacklight Configurations}

\para{Attack Configurations.} We set the L distance metrics and perturbation budgets for
different attacks following Table~\ref{table:attackconfig},
\ref{table:L2_budget}.  In these tables,  L$_\infty(x, x') =
\max_{i}(|x_i - x_i'|)$ and normalized L$_2$ distance, i.e.,   {\em
  normalized} L$_2(x, x') =
  \sqrt{\frac{1}{|x|}\sum_{i=0}^{|x|}(x_i - x_i')^2}$.

\para{Blacklight Configurations.} We list the default parameter configurations
for Blacklight in Table~\ref{table:configuration}. We discuss the impact of those
parameters in \S\ref{sec:param_analysis} and \S\ref{sec:theory_app}.
we include Blacklight configurations for both 4 image classification tasks and
the text classification task we use in \S\ref{sec:other_domain}.

\begin{table}[t]
 \centering
\resizebox{.48\textwidth}{!}{
\begin{tabular}{l|ll||l|ll}
\hline
\begin{tabular}[c]{@{}l@{}}Attack\end{tabular} &
             \begin{tabular}[c]{@{}l@{}}Distance \\
                Metric\end{tabular}
           &
             \begin{tabular}[c]{@{}l@{}}Perturbation \\
                Budget\end{tabular}
& \begin{tabular}[c]{@{}l@{}}Attack\end{tabular} &
             \begin{tabular}[c]{@{}l@{}}Distance \\
                Metric\end{tabular}
           &
             \begin{tabular}[c]{@{}l@{}}Perturbation \\
                Budget\end{tabular} \\
 \hline
 NES - QL   & L$_\infty$      & \begin{tabular}[c]{@{}l@{}}0.05 \\ (0.1 for \mnist)\end{tabular}  &
 NES - LO   &L$_\infty$      & \begin{tabular}[c]{@{}l@{}}0.05  \\(0.1 for \mnist)\end{tabular}    \\ \hline
 Boundary           & {\em normalized} L$_2$             & 0.05  &
 ECO                   & L$_\infty$        & \begin{tabular}[c]{@{}l@{}}0.05 \\ (0.1 for \mnist)\end{tabular}     \\ \hline
 HSJA                 & {\em normalized} L$_2$               & 0.05  &
 QEBA                 & {\em normalized} L$_2$               & 0.05  \\ \hline
 SurFree                 & {\em normalized} L$_2$               & 0.05  &
 Policy-Driven                 & {\em normalized} L$_2$               & 0.05  \\ \hline
\end{tabular}}

\caption{\em Black-box attack configurations. For brevity, we report the {\em normalized} L$_2$ distance
in the Perturbation Budget for L$_2$ distance metric since L$_2$ distance varies a lot according to input sizes.
We report the corresponding L$_2$ distance for different tasks in
Table~\ref{table:L2_budget}. }
\label{table:attackconfig}
\end{table}

\begin{table}
 \resizebox{.48\textwidth}{!}{
   \begin{tabular}{c|cc||c|cc}
   \hline
       Task     & {\em normalized} L$_2$ & L$_2$ & Task     & {\em normalized} L$_2$ & L$_2$   \\ \hline
   \mnist    & 0.05          & 1.4  & \gtsrb    & 0.05          & 4.2  \\
   \cifar    & 0.05          & 2.8  & \imagenet & 0.05          & 19.4 \\ \hline
 \end{tabular}}
   \caption{\em The {\em normalized} L$_2$ distance and corresponding L$_2$ distance budgets
   for different tasks we use in our experiments.}
   \label{table:L2_budget}
\end{table}

\begin{table}[t]
   \centering
 \resizebox{0.45\textwidth}{!}{
 \begin{tabular}{l|cccc|c}
\hline
\multirow{2}{*}{Task}                                                      & \multicolumn{4}{c|}{Image classification}                                                       & Text classification \\ \cline{2-6}
                                                                           & \multicolumn{1}{c}{\mnist} & \multicolumn{1}{c}{\gtsrb} & \multicolumn{1}{c}{\cifar} & \imagenet & \imdb                \\ \hline
\begin{tabular}[c]{@{}l@{}}Quantization \\ step ($\mathbf{q}$)\end{tabular}           & \multicolumn{1}{c|}{50}    & \multicolumn{1}{c|}{50}    & \multicolumn{1}{c|}{50}    & 50       & 50                  \\
\begin{tabular}[c]{@{}l@{}}Sliding window \\ size ($\mathbf{w}$)\end{tabular}         & \multicolumn{1}{c|}{50}    & \multicolumn{1}{c|}{20}    & \multicolumn{1}{c|}{20}    & 50       & 10                  \\
Sliding step ($\mathbf{p}$)                                                           & \multicolumn{1}{c|}{1}     & \multicolumn{1}{c|}{1}     & \multicolumn{1}{c|}{1}     & 1        & 1                   \\
\begin{tabular}[c]{@{}l@{}}\# of hashes per\\ fingerprint ($\mathbf{S}$)\end{tabular} & \multicolumn{1}{c|}{50}    & \multicolumn{1}{c|}{50}    & \multicolumn{1}{c|}{50}    & 50       & 30                  \\
\begin{tabular}[c]{@{}l@{}}Matching \\ threshold ($\mathbf{T}$)\end{tabular}          & \multicolumn{1}{c|}{25}    & \multicolumn{1}{c|}{25}    & \multicolumn{1}{c|}{25}    & 25       & 15                  \\ \hline
\end{tabular}}
\caption{\em Experiment configuration of Blacklight.}
 \label{table:configuration}
\end{table}

\section{Additional Results for \S\ref{sec:eval} Evaluation}
\label{sec:eval_app}
We now present additional results for \S\ref{sec:eval} including Blacklight performance
on boundary attacks with 1 million query limits, Blacklight performance on universal
patch attacks, and the detailed results for Blacklight parameter configuration impacts.

\para{Boundary attacks with 1 million queries.} Table~\ref{table:boundary_results} shows that
blacklight still has 100\% attack detect rates for boundary attacks with 1 million query limits.
Furthermore, we find that the detection coverages are even higher for attacks with 1 million query limits
than those with 100K query limits. This validates our hypothesis that Blacklight detects boundary attacks
at later stage because boundary attack advances slower in converging to the successful adversarial examples.
Finally, boundary attacks still have 0\% attack success rate with Blacklight mitigation even with 1 million
queries.

  \begin{table}[t]
    \centering
    \resizebox{.48\textwidth}{!}{
    \begin{tabular}{c|ccc|c|cc}
        \hline
    \multirow{2}{*}{Task} &\multicolumn{3}{c|}{w. Detection} &    \multicolumn{1}{c|}{w. Mitigation}                                                                                                                                                                                                              & \multicolumn{2}{c}{w/o Blacklight}                                                                                       \\\cline{2-7}
                            & \begin{tabular}[c]{@{}l@{}}Attack\\detect \%\end{tabular} & \begin{tabular}[c]{@{}l@{}}Detection\\coverage\end{tabular} & \begin{tabular}[c]{@{}l@{}}Avg queries\\to detect\end{tabular} & \begin{tabular}[c]{@{}l@{}}Attack\\success\end{tabular} & \begin{tabular}[c]{@{}l@{}}Attack\\success\end{tabular} & \begin{tabular}[c]{@{}l@{}}Avg \# attack\\ queries\end{tabular}  \\
               \hline

                 \mnist & 100\% & 76.3\% & 16 & 0\% &  26\% & 892350\\
                 \gtsrb  & 100\% & 71.2\% & 19 & 0\% & 40\% & 902931\\
                 \cifar  & 100\% & 69.7\% & 27 & 0\% & 96\% & 829124 \\
                 \imagenet & 100\% & 97.2\% & 39 & 0\% & 79\%  & 738452 \\
                 \hline
  \end{tabular}}
   \caption{\em Blacklight's detection and mitigation
       results on Boundary attack. We stop the boundary attack if it is no successful after 1 million attack queries.
     }
     \label{table:boundary_results}
  \end{table}

\para{Blacklight’s performance on universal patch attack.} Table~\ref{tab:sparse-rs}
lists the detailed results for Blacklight's detection and mitigation results on
Sparse-RS universal patch attack.

\begin{table}[t]
  \centering
  \resizebox{.5\textwidth}{!}{
  \begin{tabular}{c|ccc|c|cc}
      \hline
  \multirow{2}{*}{Task} &\multicolumn{3}{c|}{w. Detection} &    \multicolumn{1}{c|}{w. Mitigation}                                                                                                                                                                                                              & \multicolumn{2}{c}{w/o Blacklight}                                                                                       \\\cline{2-7}
                          & \begin{tabular}[c]{@{}l@{}}Attack\\detect \%\end{tabular} & \begin{tabular}[c]{@{}l@{}}Detection\\coverage\end{tabular} & \begin{tabular}[c]{@{}l@{}}Avg queries\\to detect\end{tabular} & \begin{tabular}[c]{@{}l@{}}Attack\\success\end{tabular} & \begin{tabular}[c]{@{}l@{}}Attack\\success\end{tabular} & \begin{tabular}[c]{@{}l@{}}Avg \# attack\\ queries\end{tabular}  \\
             \hline

               \mnist & 100\% & 98.4\% & 8 & 0\% & 32.9\%  & 88021\\
               \gtsrb  & 100\% & 98.9\% & 14 & 0\% & 10.8\%  & 98386\\
               \cifar  & 100\% & 97.6\% & 12 & 0\% & 54.7\%  & 87201\\
               \imagenet  & 100\% & 98.7\% & 9 & 0\% & 27.7\%  & 92039\\
               \hline
\end{tabular}
}
\caption{\em Blacklight’s detection and mitigation results on Sparse-RS universal patch attack. }
\label{tab:sparse-rs}
\end{table}

\begin{figure*}[t]
   \centering
   \includegraphics[width=.95\textwidth]{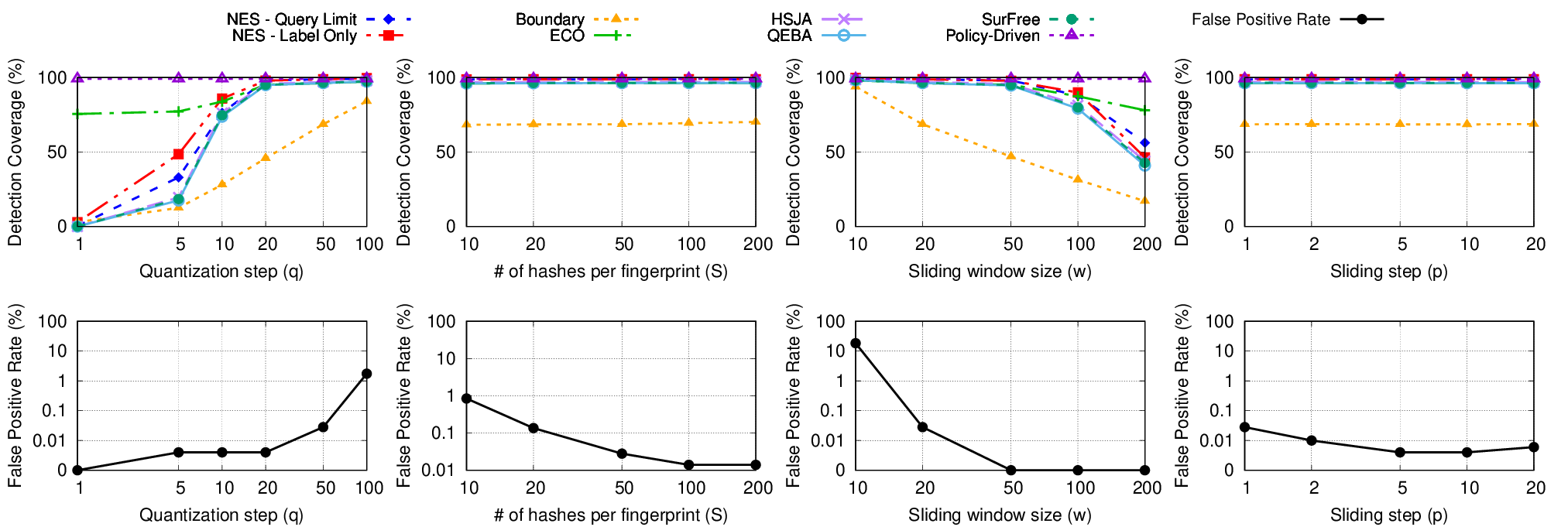}
     \vspace{-0.05in}
     \caption{\em Detection Coverage (\%) and False Positive Rate (\%) with different settings on Blacklight parameters:
     Quantization step ($\mathbf{q}$), \# of hashes per fingerprint ($\mathbf{S}$), Sliding window size ($\mathbf{w}$), and Sliding step ($\mathbf{p}$).}
   \label{fig:parameter_analysis}
 \end{figure*}

\para{Impacts for Blacklight parameter configuration.} We show the experimental
results for the impact of Blacklight parameters (Quantization step ($\mathbf{q}$), \# of hashes per fingerprint ($\mathbf{S}$), Sliding window size ($\mathbf{w}$), and Sliding step ($\mathbf{p}$)) by plotting the Detection Coverage (\%) and False Positive Rate (\%)
with different parameter settings in Figure~\ref{fig:parameter_analysis}.

\section{Additional Results for \S\ref{sec:countermeasure} Adaptive Attacks}
\label{sec:additional_results}

We now provide more analysis on countermeasures.



  \subsection{Evasion via Image Transformations.}
\label{sec:additional_results_image_trans}
  We report the details for
  our experiments against Image Transformations here. After
  applying these transformations to attack queries, we report the attack
  success rate (without the Blacklight defense) and Blacklight's attack
  detection rate, on the \cifar{} task.  Like before, we report attack detection
  rate only {\em successful attacks}.  For each setting, we run 20
  attack instances.

  For Gaussian noise based transformations, we vary the standard
  deviation (STD) of noise from 0.0001 to 0.05 (with all query inputs
  normalized to [0,1]).  Results in Table~\ref{tab:transformation} show that as noise levels
  increase, attack success rates drop quickly.   But at all noise levels tested,
  Blacklight is able to detect all successful attacks.  Intuitively,
  sufficiently high noise will disrupt classification of both benign and
  attack queries, thus degrading the attack success rate.   We see that
  Blacklight is generally more robust than the
  attack's iterative optimization process -- Blacklight continues to detect attacks at noise levels
  where the noise has long since disrupted the attack.

  For image augmentation, we test 4 cases where the attacker shifts
  each input horizontally/vertically by up to 10\%, rotate by up to
  $10^{\circ}$, zoom in by up to 10\%, and a combination of all three.
  Table~\ref{tab:transformation}
  shows that while different attacks react differently to image augmentation
  techniques (some still produce successful attacks while others fail
  completely), Blacklight is able to detect all successful attack sequences under
  different transformations.

  \begin{table}[t]
    \centering
    \resizebox{.5\textwidth}{!}{
    \begin{tabular}{l|l|cccc|cccc}
    \hline
    \multicolumn{2}{l|}{\multirow{2}{*}{\backslashbox[35mm]{\bf Attack}{\bf Transformation}}} & \multicolumn{4}{c|}{\bf Gaussian Noise w. Different STD} & \multicolumn{4}{c}{\bf Image Augmentation} \\ \cline{3-10}

      \multicolumn{2}{l|}{}      &\bf  0.0001    & \bf 0.0005    & \bf 0.005   & \bf 0.05    & \bf Shift   & \bf Rotate   & \bf Zoom   & \bf Comb.  \\ \hline

      \multirow{2}{*}{\bf NES - QL}       & \bf ASR       & 85\%      & 80\%      & 15\%    & 0\%     & 100\%   & 75\%       & 80\%   & 60\%      \\

  & \bf ADR                & 100\%     & 100\%     & 100\%   & N/A         & 100\%   & 100\%      & 100\%  & 100\%     \\ \hline
  \multirow{2}{*}{\bf NES - LO}                 & \bf ASR                                         & 25\%      & 20\%     & 15\%     & 0\%     & 100\%   & 45\%       & 70\%   & 20\%      \\

  &\bf  ADR                & 100\%     & 100\%    & 100\%   & N/A         & 100\%   & 100\%      & 100\%  & 100\%     \\ \hline
    \multirow{2}{*}{\bf Boundary}                         & \bf ASR                                         & 90\%      & 90\%      & 85\%    & 0\%     & 90\%    & 90\%       & 90\%   & 90\%      \\

  & \bf ADR                & 100\%     & 100\%     & 100\%   & N/A         & 100\%   & 100\%      & 100\%  & 100\%     \\ \hline
    \multirow{2}{*}{\bf ECO}                              & \bf ASR                                         & 85\%      & 0\%       & 0\%     & 0\%     & 0\%     & 0\%        & 0\%    & 0\%       \\

  & \bf ADR                & 100\%     & N/A         & N/A       & N/A         & N/A       & N/A          & N/A      & N/A         \\ \hline
    \multirow{2}{*}{\bf HSJA}                             & \bf ASR                                         & 95\%      & 20\%      &  5\%   & 0\%     &  0\%  &    5\%   &  10\%  &   15\%   \\

  &\bf  ADR                & 100\%     & 100\%      & 100\%   & N/A         &  N/A  &     100\%  &    100\% &  100\%   \\ \hline
    \end{tabular}
  }
  \caption{\em  Attack success rate (ASR) w/o Blacklight mitigation and Blacklight attack detection rate (ADR) of successful attacks
    as attackers add different image transformations. Column 3-6 report the results for adding Gaussian Noise with different standard deviation (STD) and Column 7-10
    report the results for applying different image transformations to each attack queries.}
   \vspace{-0.2in}
  \label{tab:transformation}
  \end{table}

\subsection{Increasing Perturbation Budget}
\label{sec:additional_results_increase_purb}

In order to provide a comprehensive evaluation on the impact of increasing perturbation
budget on the detection performance for Blacklight, we run experiments on all tasks for
the two fastest converging attacks (ECO and HSJA) with larger perturbation budgets.
Table~\ref{table:perturb_all_models} shows that Blacklight achieves 100\%  on all tasks
for ECO attacks even with perturbation budget up to $0.2$. For HSJA attack, Blacklight
can detect 100\% of attacks on all tasks when the {\em normalized} L$_2$ perturbation budgets
are no more than 0.1. When the {\em normalized} L$_2$ perturbation budgets get larger,
Blacklight's detection rate drops gradually. However, we believe this is reasonable
since the {\em normalized} L$_2$ budget is too large that even exceeds the {\em normalized} L$_2$
distances between some benign images.

\begin{table}[t]
  \centering
  \resizebox{.48\textwidth}{!}{
  \begin{tabular}{l|cccc|cccc}
  \hline
  \multirow{2}{*}{Task} & \multicolumn{4}{c|}{ECO} & \multicolumn{4}{c}{HSJA} \\ \cline{2-9}
                                                                          & 0.05        & 0.1         & 0.15       & 0.2        & 0.05        & 0.1        & 0.15       & 0.2        \\ \hline
  \mnist                                                                & 100\%       & 100\%       & 100\%      & 100\%      & 100\%       & 100\%      & 75\%      & 40\%      \\
  \gtsrb                                                                & 100\%       & 100\%       & 100\%      & 100\%      & 100\%       & 100\%      & 70\%      & 50\%      \\
  \cifar                                                                & 100\%       & 100\%       & 100\%       & 100\%       & 100\%       & 100\%      & 80\%      & 40\%       \\
  \imagenet                                                                     & 100\%       & 100\%       & 100\%      & 100\%      & 100\%       & 100\%      & 90\%      & 85\%      \\ \hline
  \end{tabular}}
\caption{\small \em Blacklight detection rate for attacks using larger perturbation
  budgets for all tasks. We use $\mathbf{T}=15$ with a small increase in false
  positives (0.74\%).}
\label{table:perturb_all_models}
\end{table}

\begin{figure*}[t]
  \centering
  \includegraphics[width=\textwidth]{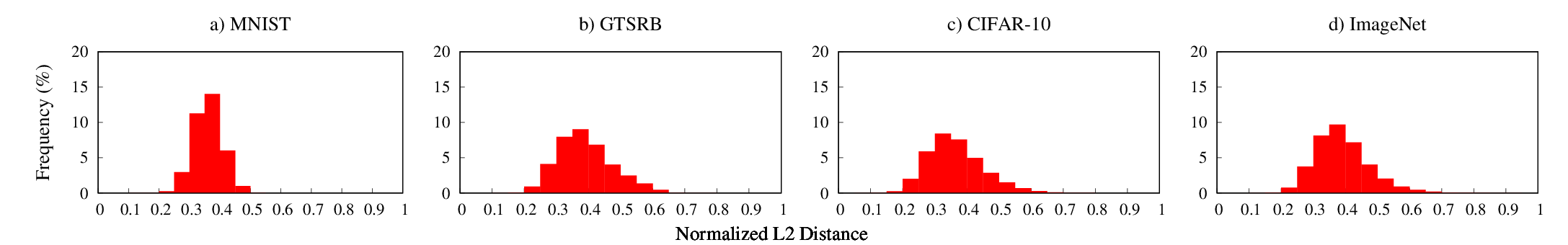}
    \vspace{-0.05in}
    \caption{\em Frequency of the {\em normalized} $L_2$ distances between benign images from different labels for all tasks.}
  \label{fig:l2_baseline}
\end{figure*}

\begin{figure}[t]
  \centering
  \includegraphics[width=.5\textwidth]{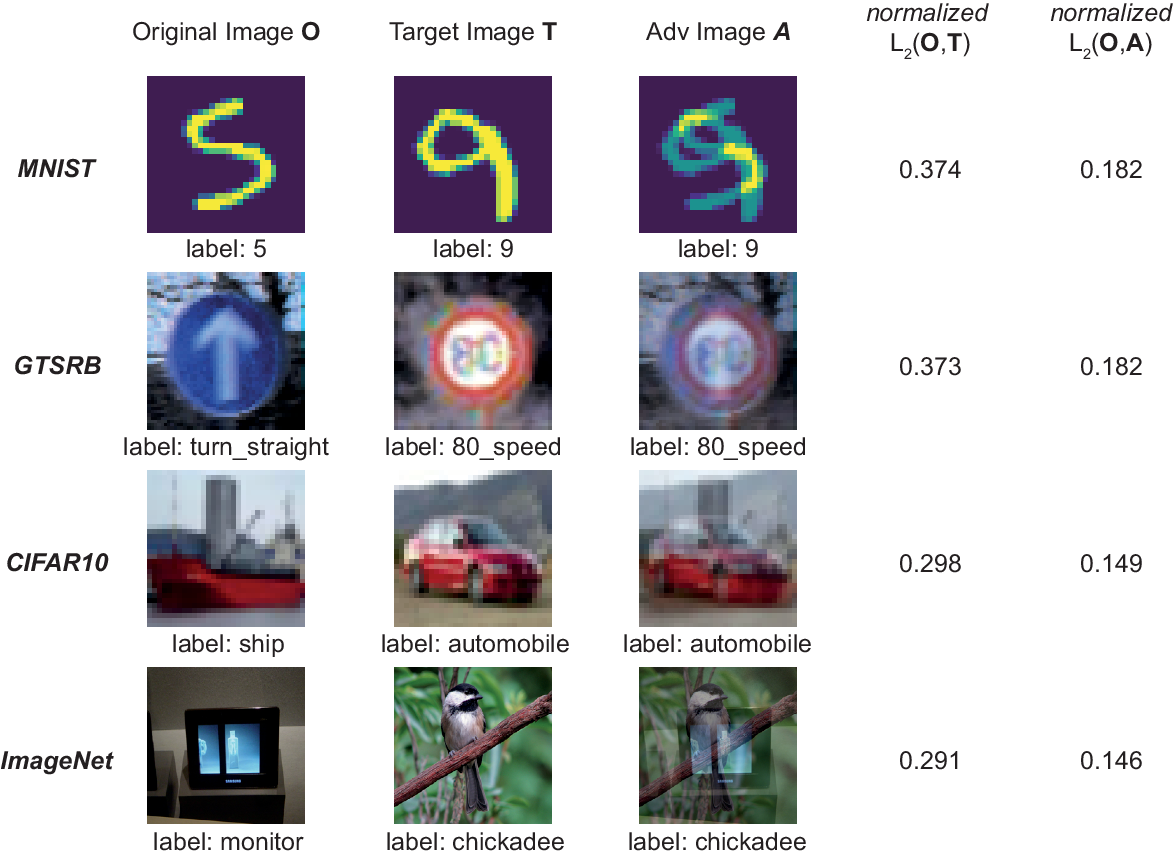}
    \vspace{-0.05in}
    \caption{\em Examples of successful adversarial attacks via blending two benign images when the perturbation
    budget is set to $0.2$.}
  \label{fig:l2_example}
\end{figure}

We analyze the distribution of the {\em normalized} $L_2$ distances between benign images from
different labels. We randomly pick 50K benign image pairs from different labels and calculate the
{\em normalized} $L_2$ distances between these pairs. Figure~\ref{fig:l2_baseline} shows the
distribution for the {\em normalized} $L_2$ distances between benign images from different labels for all
four tasks. We can see that for all tasks, the majority of image pairs have a {\em normalized} $L_2$ distance
no more than $0.4$ and there are a significant proportion of benign image pairs having a {\em normalized} $L_2$ distance
no more than $0.3$.

We say that a reasonable $L_2$ perturbation budget for adversarial examples should
smaller than half of the {\em normalized} $L_2$ distances for most of the benign image pairs. Otherwise,
by simply blending two benign images (calculating the mean of two images), the attacker can create a
successful adversarial example: assume the attacker has an image pair $(x_a, x_b)$, the model $\mathbb{F}$ classifies $x_a$ to
label $A$ and $x_b$ to label $B$, $x' = \frac{x_a+x_b}{2}$ cannot be classified both to label $A$ and $B$. If {\em normalized} $L_2(x_a, x')$
is smaller than perturbation budget, $x'$ is an adversarial example for target label $\mathbb{F}(x')$ either from original
image $x_a$ or from image $x_b$.

Figure~\ref{fig:l2_example} shows examples where the attacker can successfully create an adversarial
example by simply calculating the average of two benign images when the {\em normalized} $L_2$ budget is
set to $0.2$. In such case, although the attack will succeed within one single query, we believe this is not a reasonable
perturbation budget for adversarial attacks.

\subsection{Guided Transformations when Attacker Knows ($\mathbf{q},\mathbf{p},\mathbf{w}$).}
We show the algorithm we use for guided transformations when attacker knows ($\mathbf{q},\mathbf{p},\mathbf{w}$)
in Algorithm~\ref{alg:guided_trans_alg}.

\begin{algorithm}
  \caption{Algorithm for Guided Transformation Attacks when Attacker Knows ($\mathbf{q},\mathbf{p},\mathbf{w}$)}
  \label{alg:guided_trans_alg}
    \hspace*{\algorithmicindent} \textbf{Parameter:} Sliding window size of Blacklight: $\mathbf{w}$, quantization step of Blacklight: $\mathbf{q}$\\
    \hspace*{\algorithmicindent} \textbf{Input:} Attack query $x$\\
    \hspace*{\algorithmicindent} \textbf{Output:} Modified attack query $x$
\begin{algorithmic}[1]
  \Procedure{Initization}{$\mathbf{w}$, $\mathbf{q}$}
  \State \# We save all combinations for pixels modification in a queue.
  \State $PermList \gets [ ]$
  \For{$i = 1\, \textbf{ to }\, \mathbf{w}$}
  \State \# compute all combinations for selecting $i$ pixels from $\mathbf{w}$ pixels, which generates $C_{\mathbf{w}}^{i}$ choices.
  \State pixelCombination = Combination($i$, $w$)
  \State \# For each pixel selected there are 2 choices for combinations ($+\mathbf{q}/-\mathbf{q}$), which generates $2^i \times C_{\mathbf{w}}^{i}$ choices in total.
  \State allPixelCombination = Update2ChoicesPerPixel (pixelCombination)
  \State PermList.append(allPixelCombination)
\EndFor
\State \Return PermList
\EndProcedure

\Procedure{GuidedTransformation}{$x$}
  \State \# we pop the first element from the queue, which is the modification choice with smallest number of pixel changes in the remaining choices.
  \State CurrentPermutation = PermList.pop()
  \State Apply the modification for CurrentPermutation to every $\mathbf{w}$ pixels of $x$.
\State \Return x
\EndProcedure
\end{algorithmic}
\end{algorithm}

%

\subsection{Optimal Black-Box Attacks.}
\label{sec:opt}
We give more discussion in optimal black-box attacks.
First, to simulate a near-optimal query-efficient attack, we evenly downsample
attack query sequences from 5 attacks to generate attack
sequences that are a tiny fraction of current sequences. We then test Blacklight's
detection performance on these subsampled attack sequences.
Table~\ref{table:stronger_attack} shows that even when attacks are able to
complete in 500, 100, or 50 queries, Blacklight still detects them near
perfectly (100\% detection for 4 attacks and 89\% for Boundary attack). Even
when these attacks complete within 10 queries, Blacklight is still highly
successful at detecting NES-QL, ECO and HSJA.

We note that NES-LO and Boundary attacks have much lower detection rates than
other attacks when only choosing 10 queries from attack sequences. This
is because both NES-LO and Boundary attacks are both boundary attacks that
jump back and forth between two images
(original and target image).  Random subsets of 10 out of
thousands of queries are more likely to be variants of the source or target
that are sufficiently different from each other as to avoid detection.

Second, for ``perfect-gradient'' black-box algorithm, each iteration of the
gradient calculation for an analogous white-box attack would translate to a
single query over the network by the black-box attacker.  This idealized
black-box attack uses CW~\cite{carlini2017towards} and
PGD~\cite{madry2017towards} to generate attack sequences against our \cifar{}
model.  On average, CW and PGD converge after only $6.3$ and $3.1$
queries. Against simulated black-box attacks using these attack queries,
Blacklight detects $100\%$ of attacks driven by CW, and $81\%$ of PGD-driven
attacks.

\begin{table}[t]
  \centering
  \resizebox{.45\textwidth}{!}{
\begin{tabular}{l|c|c|c|c}
\hline
\backslashbox{Attack Type}{$N$}       & 500               & 100              & 50               & 10               \\ \hline
NES - Query Limit & 100\%             & 100\%            & 100\%            & 95\%             \\
NES - Label Only  & 100\%             & 100\%            & 100\%            & 31\%             \\
Boundary          & 100\%             & 90\%             & 89\%             & 48\%             \\
ECO               & 100\%             & 100\%            & 100\%            & 100\%            \\
HSJA              & 100\%             & 100\%            & 100\%            & 91\%             \\ \hline
CW                & \multicolumn{4}{l}{Average $N$ = 6.33, Detection rate = 100\%} \\
PGD               & \multicolumn{4}{l}{Average $N$ = 3.13, Detection rate = 81\%}  \\ \hline
\end{tabular}}

\caption{\em Blacklight's performance against near-optimal
  ``query-efficient'' and ``perfect-gradient'' black-box attacks.}
\label{table:stronger_attack}
\end{table}

\subsection{Pause and Resume Attacks.} Table~\ref{table:cycles} shows the exact number
of average reset cycles needed for different attacks. We also include average total
queries needed for attacks as reference.

\begin{table}[t]
  \centering
  \resizebox{.44\textwidth}{!}{
  \begin{tabular}{@{\extracolsep{1pt}}l|c|c}
      \hline
      Attack Type & Average Reset Cycles Needed & Average Total Queries\\
      \hline
      NES-QL & 11471 & 12695 \\
      NES-LO  & 65837 & 67099 \\
      Boundary   & 2285 & 6160 \\
      ECO             & 16590 & 16591 \\
      HSJA            & 1092 & 1121 \\
      \hline
  \end{tabular}}
\caption{\em Average reset cycles needed for a successful Pause and Resume attack
  on \cifar{}. The fastest attack (HSJA) can succeed in roughly 3 years.}
\label{table:cycles}
\end{table}

\end{document}